\documentclass[12pt,preprint]{aastex}

\begin{document}

\title{Modeling the Evolution and Distribution of the Frequency's Second Derivative and Braking Index of Pulsar Spin with Simulations}
\author{Yi Xie\altaffilmark{1,2}, Shuang-Nan Zhang\altaffilmark{1,3}}
\altaffiltext{1}{National Astronomical Observatories, Chinese
Academy of Sciences, Beijing, 100012, China}
\altaffiltext{2}{University of Chinese Academy of Sciences, Beijing,
100049, China}
\altaffiltext{3}{Key Laboratory of Particle Astrophysics, Institute
of High Energy Physics, Chinese Academy of Sciences, Beijing 100049,
China; zhangsn@ihep.ac.cn}

\begin{abstract}
We model the evolution of spin frequency's second derivative
$\ddot\nu$ and braking index $n$ of radio pulsars with simulations
within the phenomenological model of their surface magnetic field
evolution, which contains a long-term decay modulated by short-term
oscillations. For the pulsar PSR B0329+54, the model can reproduce
the main characteristics of its $\ddot\nu$ variation with
oscillation periods, predicts another $\sim 50$ yr oscillation component and another recent swing of the sign of $\ddot\nu$. We show that the ``averaged'' $n$ is different
from the instantaneous $n$, and its oscillation magnitude decreases
abruptly as the time span increases, due to the ``averaging''
effect. The simulation predicted timing residuals agree with the
main features of the reported data. We further perform Monte Carlo
simulations for the distribution of the reported data in
$|\ddot\nu|$ versus characteristic age $\tau_{\rm c}$ diagram. The
model with a power law index $\alpha=0.5$ can reproduce the slope of
the linear fit to pulsars' distributions in the diagrams of
$\log|\ddot\nu|-\log\tau_{\rm c}$ and $\log|n|-\log\tau_{\rm c}$,
but the oscillations are responsible for the almost equal number of
positive and negative values of $\ddot\nu$, in agreement with our
previous analytical studies; an oscillation period of about several
decades is also preferred. However the range of the oscillation
amplitudes is $-11.4\lesssim\log f\lesssim-10.2$, slightly lager
than the analytical prediction, $\log f\simeq-11.85$, because the
``averaging'' effect was not included previously.
\end{abstract}


\section{Introduction}    
\label{sect:intro}

The spin-down of radio pulsars is caused by emitting electromagnetic
radiation and by accelerating particle winds. Traditionally, the
evolution of their rotation frequencies $\nu$ may be described by
the braking law
\begin{equation}\label{braking law}
\dot\nu =-K \nu^n,
\end{equation}
where $n$ is the braking index, $K$ is a positive constant that
depends on the magnetic dipole moment and the moment of inertia of
the neutron star. By differentiating Equation (\ref{braking law}),
one can obtain $n$ in terms of observables, $ n=\ddot\nu
\nu/\dot\nu^2$. For the standard vacuum magnetic dipole radiation
model with constant magnetic fields (i.e. $\dot K=0$), $n=3$
(Manchester \& Taylor 1977). Thus the frequency's second derivative
can be simply expressed as
\begin{equation}\label{ddotnu}
\ddot\nu=3\dot\nu^2/\nu.
\end{equation}
The model predicts $\ddot\nu>0$ and $|\ddot\nu|$ should be very
small.

However, unexpectedly large values of $\ddot\nu$ were measured for
several dozen pulsars thirty years ago (Gullahorn \& Rankin 1978;
Helfand et al. 1980; Manchester \& Taylor 1977), and many of those
pulsars surprisingly showed $\ddot\nu<0$. Some authors suggested
that the observed values of $\ddot\nu$ could result from a
noise-type fluctuation in the pulsar period (Helfand et al. 1980;
Cordes 1980; Cordes \& Helfand 1980). Based on the timing data of
PSR B0329+54, Demia${\rm \acute{n}}$ski \& Pr${\rm
\acute{o}}$szy${\rm \acute{n}}$ski (1979) further proposed that a
distant planet would influence $\ddot\nu$, and the quasi-sinusoidal
modulation in timing residuals might be caused by changes in pulse
shape, precession of a magnetic dipole axis, or an orbiting planet.
Baykal et al. (1999) investigated the stability of $\ddot\nu$ of PSR
B0823+26, B1706-16, B1749-28 and B2021+51 using their
time-of-arrival (TOA) data extending to more than three decades,
confirmed that the anomalous $\ddot\nu$ terms of these sources
arise from red noise (timing residual with low frequency
structure), which may originate from the external torques from the
magnetosphere of a pulsar.

The relationship between the low frequency structure in timing
residuals and the fluctuations in pulsar spin parameters ($\nu$,
$\dot\nu$, and $\ddot\nu$) is very interesting and important. We
call both the residuals and the fluctuations as the ``timing noise''
in the present work, since we will infer that they have the same
origin. Timing noise for some pulsars has even been studied over
four decades (e.g. Boynton et al. 1972; Groth 1975; Jones 1982;
Cordes \& Downs 1985; D'Alessandro et al. 1995; Kaspi, Chakrabarty
\& Steinberger 1999; Chukwude 2003; Livingstone et al. 2005; Shannon
\& Cordes 2010; Liu et al. 2011; Coles et al. 2011; Jones 2012).
However, the origins of the timing noise are still controversial and
there is still unmodelled physics to be understood. Boynton et al.
(1972) suggested that the timing noise might arise from ``random
walk'' processes. The random walk in $\nu$ may be produced by small
scale internal superfluid vortex unpinning (Alpar, Nandkumar \&
Pines 1986; Cheng 1987a), or short time ($t\sim 10~{\rm ms}$ for
Crab pulsar) fluctuations in the size of outer magnetosphere gap
(Cheng 1987b). Stairs, Lyne \& Schemar (2000) reported long
time-scale, highly periodic and correlated variations in the pulse
shape and the slow-down rate of the pulsar PSR B1828-11, which have
generally been considered as evidence of free precession. The
possibilities were also proposed that the quasi-periodic modulations
in timing residuals could be caused by an orbiting asteroid belt
(Cordes \& Shannon 2008) or a fossil accretion disk (Qiao et al.
2003).

Recently, Hobbs et al. (2010, hereafer H2010) carried out so far the
most extensive study of the long term timing irregularies of 366
pulsars. Besides ruling out some timing noise models in terms of
observational imperfections, random walks, and planetary companions,
some of their main conclusions are: (1) timing noise is widespread
in pulsars and is inversely correlated with pulsar characteristic
age $\tau_{\rm c}$; (2) significant periodicities are seen in the
timing noise of a few pulsars, but quasi-periodic features are
widely observed; (3) the structures seen in the timing noise vary
with data span, i.e., more quasi-period features are seen for longer
data span and the magnitude of $|\ddot{\nu}|$ for shorter data span
is much larger than that caused by magnetic braking of the neutron
star; and (4) the numbers of negative and positive $\ddot{\nu}$ are
almost equal in the sample, i.e. $N(\ddot\nu>0)\thickapprox
N(\ddot\nu<0)$. Lyne et al. (2010) showed credible evidence that
timing noise and $\dot\nu$ are correlated with changes in the pulse
shapes, and are therefore linked and caused by the changes in the
pulsar's magnetosphere.

Blandford \& Romani (1988) re-formulated the braking law of a pulsar
as $\dot\nu=-K(t)\nu^3$, which means that the standard magnetic
dipole radiation is still responsible for the instantaneous
spin-down of a pulsar, and $\ddot\nu\nu/\dot\nu^2\neq3$ does not
indicate deviation from the dipole radiation model, but means only
that $K(t)$ is time dependent. Considering the magnetospheric origin
of timing noise as inferred by Lyne et al. (2010), we assume that
magnetic field evolution is responsible for the variation of $K(t)$,
i.e. $K=A B(t)^2$, in which $A=\frac{8\pi^2R^6\sin\theta^2}{3c^3I}$
is a constant, $R~(\simeq10^6~{\rm cm})$, $I~(\simeq10^{45}~{\rm
g~cm^2})$, and $\theta~(\simeq\pi/2)$ is the radius, moment of
inertia, and angle of magnetic inclination of the neutron star,
respectively. We can rewrite Equation (\ref{ddotnu}) as
\begin{equation}\label{ddotnu-2}
\ddot\nu=3\dot\nu^2/\nu+2\dot\nu\dot B/B.
\end{equation}
Since the numbers of negative and positive $\ddot{\nu}$ are almost
equal, it should be the case that $B$ quasi-symmetrically
oscillates, and usually $|2\dot\nu\dot B/B|\gg3\dot\nu^2/\nu$.
Meanwhile, it is noticed that pulsars with $\tau_{\rm
c}\lesssim10^5~{\rm yr}$ always have $\ddot\nu>3\dot\nu^2/\nu$
(H2010); a reasonable understanding is that their magnetic field
decays (i.e. $\dot B<0$) dominate the field evolution for these
``young'' pulsars.

Therefore, Zhang \& Xie (2012a, hereafter Paper I) constructed a
phenomenological model for the dipole magnetic field evolution of
pulsars with a long-term decay modulated by short-term oscillations,
\begin{equation}\label{B evolution}
B(t)=B_d(t)(1+\sum k_i\sin(\phi_i+2\pi\frac{t}{T_i})),
\end{equation}
where $t$ is the pulsar's age, and $k_i$, $\phi_i$, $T_i$ are the
amplitude, phase and period of the $i$-th oscillating component,
respectively. $B_d(t)=B_0(t/t_0)^{-\alpha}$, in which $B_0$ is the
field strength at the age $t_0$, and the index $\alpha\gtrsim0.5$
(see Paper I for details). Substituting Equation (\ref{B evolution})
into Equation (\ref{braking law}), we get the differential equation
describing the the spin frequency evolution of a pulsar as follows
\begin{equation}\label{braking law2}
\dot\nu \nu^{-3}=-A B(t)^2.
\end{equation}

In paper I, we showed that the distribution of $\ddot\nu$ and the
inverse correlation of $\ddot\nu$ versus $\tau_c$ could be well
explained with analytic formulae derived from the phenomenological
model. In Zhang \& Xie (2012b, hereafter Paper II), we also derived
an analytical expression for the braking index ($n$) and pointed out
that the instantaneous value of $n$ of a pulsar is different from
the ``averaged'' $n$ obtained from the traditional phase-fitting
method over a certain time span. However, this ``averaging'' effect
was not included in our previous analytical studies; this work is
focused on addressing this effect.

This paper is organized as follows. In Section 2, we show that the
timescales of magnetic field oscillations are tightly connected to
the $\ddot\nu$ evolution and the quasi-periodic oscillations
appearing in the timing residuals, and the reported data of pulsar
B0329+54 are fitted. In Section 3, we perform Monte Carlo
simulations on the pulsar distribution in the $\ddot\nu-\tau_{\rm
c}$ and $n-\tau_{\rm c}$ diagram. Our results are summarized and
discussed in Section 4.

\section{Modeling the $\ddot\nu$ and $n$ Evolution and Timing Residuals of Pulsar B0329+54}

PSR B0329+54 is a bright (e.g. 1500 mJy at 400
MHz\footnote{http://www.atnf.csiro.au/people/pulsar/psrcat/}),
$0.71$ s pulsar that had been suspected to possess planetary-mass
companions (Demia${\rm \acute{n}}$ski \& Pr${\rm
\acute{o}}$szy${\rm \acute{n}}$ski 1979; Bailes, Lyne, \& Shemar
1993; Shabanova 1995). However the suspected companions have not
been confirmed and are currently considered doubtful (Cordes \&
Downs 1985; Konacki et al. 1999; H2010). Konacki et al. (1999)
suggested that the observed ephemeral periodicities of the timing
residuals for PSR B0329+54 are intrinsic to this neutron star. H2010
believed that the timing residual has a similar form to the other
pulsars in their sample. They plotted $|\ddot\nu|$ obtained from the
B0329+54 data sets with various time spans (see Figure 12 in their
paper). For data spanning $\sim 10$ yr, they measured a large and
significant $\ddot\nu$, and found that the timing residual takes the
form of a cubic polynomial. However, no cubic term was found for
data spanning more than $\sim 25$ yr, and $|\ddot\nu|$ became
significantly smaller. The reported periods of the timing
residuals for PSR B0329+54 are $1100~{\rm days}$, $2370~{\rm days}$,
and/or $16.8~{\rm years}$ (Demia${\rm \acute{n}}$ski \& Pr${\rm
\acute{o}}$szy${\rm \acute{n}}$ski 1979; Bailes, Lyne, \& Shemar
1993; Shabanova 1995). Here we neglect the two short-period
oscillation components, since they have little impact on $|\ddot\nu|$
variation due to their relative small oscillation
magnitude (Shabanova 1995).

In order to model the $\ddot\nu$ evolution for pulsar B0329+54, we
firstly obtain $\nu(t)$ by integrating the spin-down law described
as Equations (\ref{braking law2}) and (\ref{B evolution}) with
$\alpha=0.5$, and then the phase
\begin{equation}\label{phase integrate}
\Phi(t)=\int_{t_0}^{t}\nu(t'){\rm d}t'.
\end{equation}
Finally, these observable quantities, $\nu$, $\dot{\nu}$ and
$\ddot{\nu}$ can be obtained by fitting the phases to the third
order of its Taylor expansion over a time span $T_{\rm s}$,
\begin{equation}\label{phase}
\Phi(t_i) =\Phi_0 + \nu (t_i-t_0) + \frac{1}{2}\dot \nu (t_i-t_0)^2
+ \frac{1}{6}\ddot\nu (t_i-t_0)^3.
\end{equation}
We thus get $\nu$, $\dot{\nu}$ and $\ddot{\nu}$ for $T_{\rm s}$ from
fitting to Equation~(\ref{phase}), with a certain time interval of
phases $\Delta T_{\rm int}=10^6~{\rm s}$ (interval between each TOA,
i.e. $\Delta T_{\rm int}=t_{i+1}-t_{i}$).

\begin{figure}
\centering
\includegraphics[angle=0,scale=0.6]{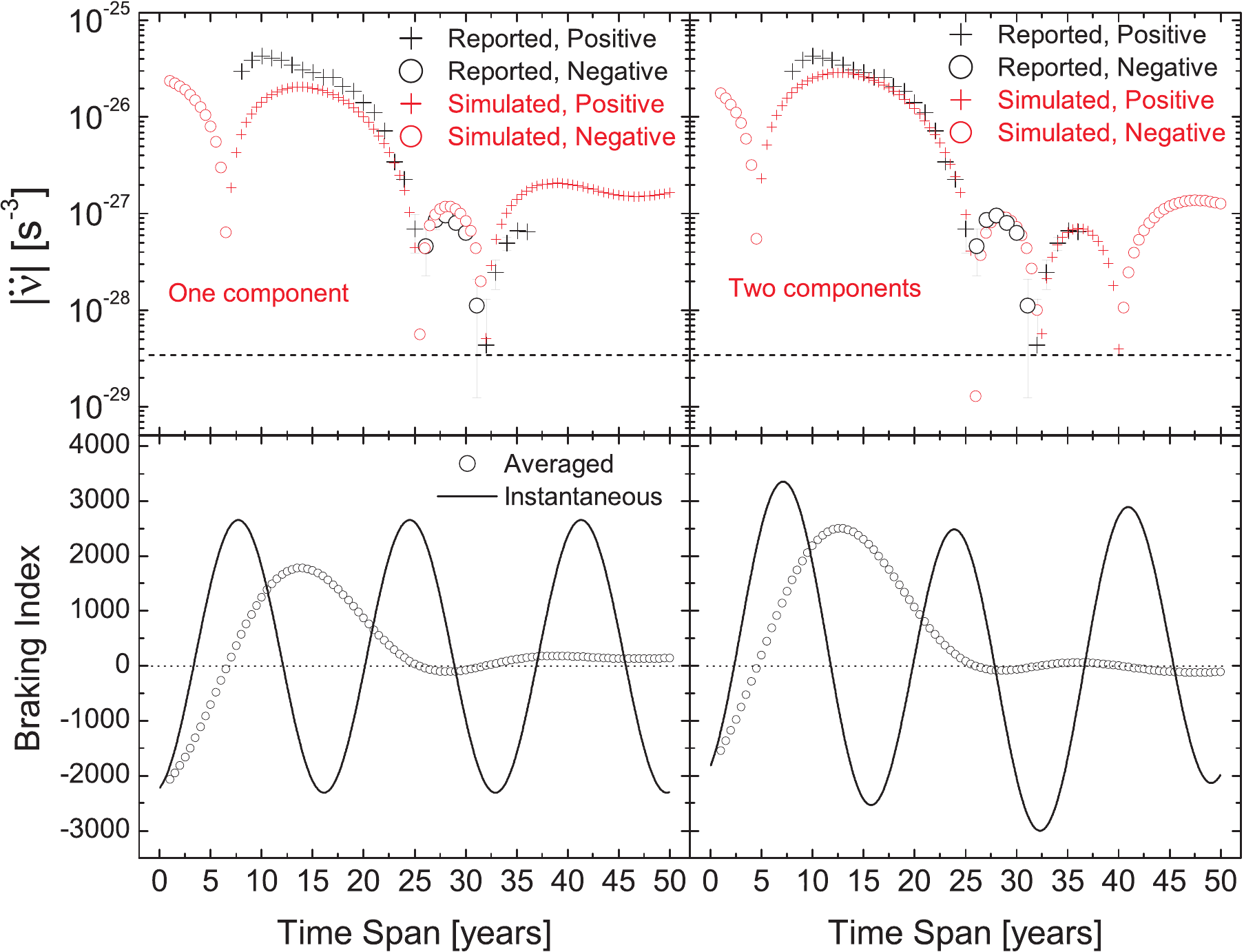}
\caption{$|\ddot\nu|$ and $n$ for PSR B0329+54. Upper panels:
$|\ddot\nu|$ obtained using different data spans ($T_s$). The values
reported by H2010 are represented by large cross symbols
($\ddot\nu>0$) and large circles ($\ddot\nu<0$); our simulated
values are represented by small cross symbols ($\ddot\nu>0$) and
small circles ($\ddot\nu<0$); the horizontal dashed line represents
$\ddot\nu=3\dot\nu^2/\nu$. Bottom panels: instantaneous (solid
lines) and averaged (circles) values of  $n$. Left panels:
simulation with one oscillation component of $k=3\times10^{-4}$,
$T=16.8~{\rm yr}$ and $\phi=3.7$. Right panels: simulation with two
oscillation components of $k_1=3.3\times10^{-4}$,
$k_2=1.8\times10^{-4}$, and $T_1=16.8~{\rm yr}$, $T_2=50~{\rm yr}$,
and $\phi_1=3.9$, $\phi_2=4.0$. The horizontal dotted line
represents $n=0$.} \label{Fig:1}
\end{figure}

In the upper panels of Figure {\ref{Fig:1}}, we show the
reported and simulated results of $|\ddot\nu|$ for various $T_{\rm
s}$ for PSR B0329+54. The reported data are read from Figure 12 of
H2010. The simulated results with one oscillation component with
fitting parameters $k=3\times10^{-4}$, $T=16.8~{\rm yr}$ and
$\phi=3.7$ are shown in the left panels, and two oscillation
components with $k_1=3.3\times10^{-4}$, $k_2=1.8\times10^{-4}$,
$T_1=16.8~{\rm yr}$, $T_2=50~{\rm yr}$, and $\phi_1=3.9$,
$\phi_2=4.0$ are shown in the right panels. Besides the $16.8$ yr
component, there is another oscillation component with period
$\sim 50$ yr in the two component model. In the bottom panels of
Figure {\ref{Fig:1}}, we show the corresponding $n$ with the same
oscillation parameters obtained above. The braking index $n=\ddot\nu
\nu/\dot\nu^2$ obtained directly from Equation (\ref{braking law2})
is called ``instantaneous'' $n$; similarly that the obtained by
fitting phase sets to Equation (\ref{phase}) is called ``averaged''
$n$. It can be seen that the averaged $n$ has the same variation
trends with $\ddot\nu$, since $|\Delta\nu/\nu| \sim10^{-6}$
and $|\Delta\dot\nu/\dot\nu| \sim10^{-3}$ are tiny, compared to
$|\Delta\ddot\nu/\ddot\nu| \sim 1$. The magnitude of the first
period of the averaged $n$ is close to the instantaneous one, but it
decays significantly due to the ``averaging'' effect.

\begin{figure}
\centering
\includegraphics[angle=0,scale=0.5]{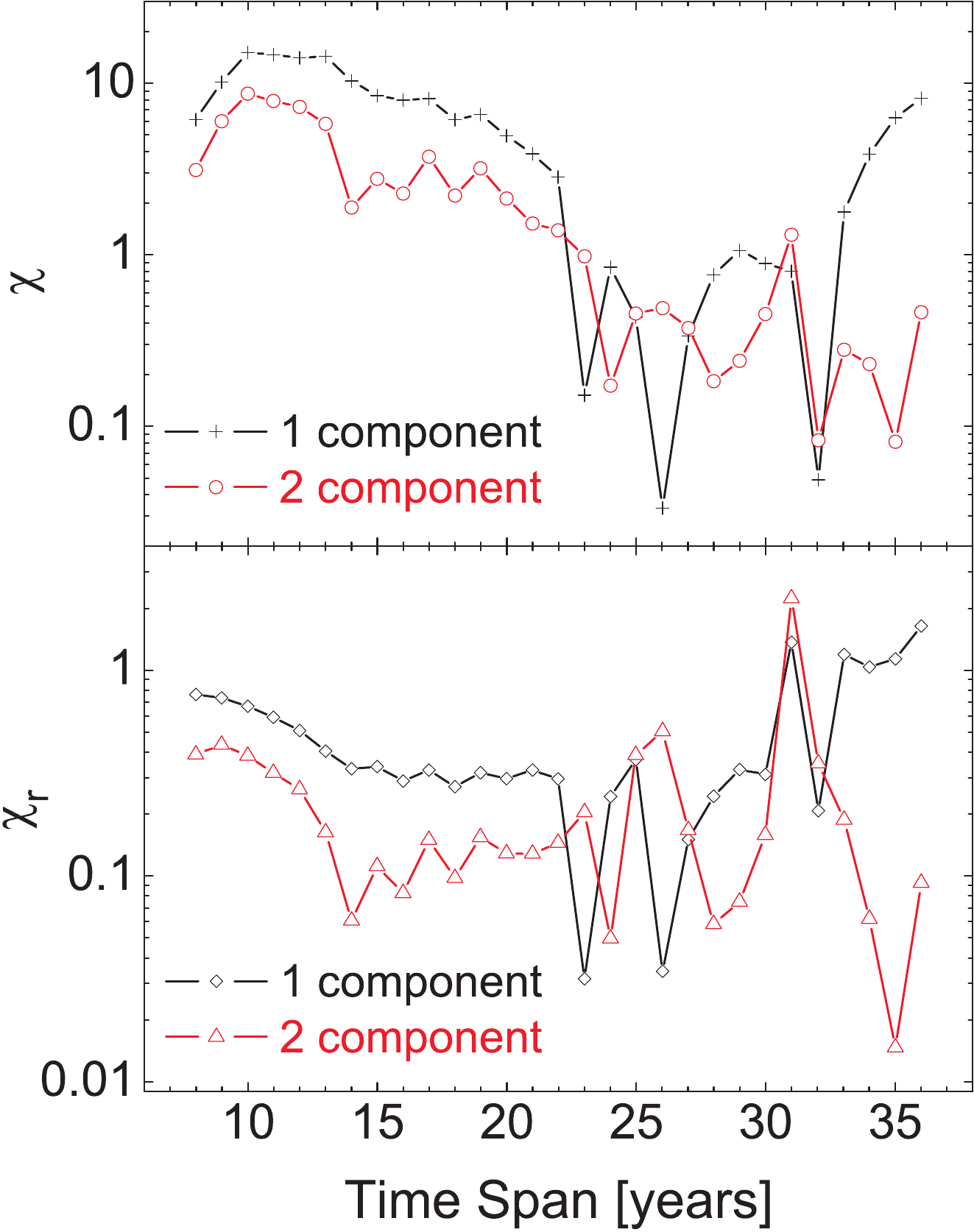}
\caption{Two goodness of fit parameters $\chi$ and $\chi_{\rm r}$
for the fits of $|\ddot\nu|$ variation. It is shown that the two
component case is apparently a better fit.} \label{Fig:r1}
\end{figure}

We adopt two goodness of fit parameters to show how well the
model matches the data, i.e. $\chi=|\frac{\ddot\nu_{\rm
M}-\ddot\nu_{\rm D}}{\sigma}|$ and $\chi_{\rm
r}=|\frac{\ddot\nu_{\rm M}-\ddot\nu_{\rm D}}{\ddot\nu_{\rm M}}|$,
where the subscripts `M' and `D' refer to the model results and the reported data, respectively, $\sigma$ the uncertainties of reported
data. $\chi$ and $\chi_{\rm r}$ are shown in Figure
{\ref{Fig:r1}}. One can see that both fits are not very good and
are certainly rejected by $\chi^2$ test. However, we stress that
both the one and two component models can reproduce the main
characteristics of $\ddot\nu$ variation, including the swings
between $\ddot\nu>0$ and $\ddot\nu<0$. On the other hand, the calculated $\chi$ and $\chi_{\rm r}$ apparently indicate that the two
component case is a better fit. Unfortunately, we cannot provide any
physical information about the links between the identified
periodicities, since the physical processes of the oscillations are
poorly understood presently.

\begin{figure}
\centering
\includegraphics[angle=0,scale=0.5]{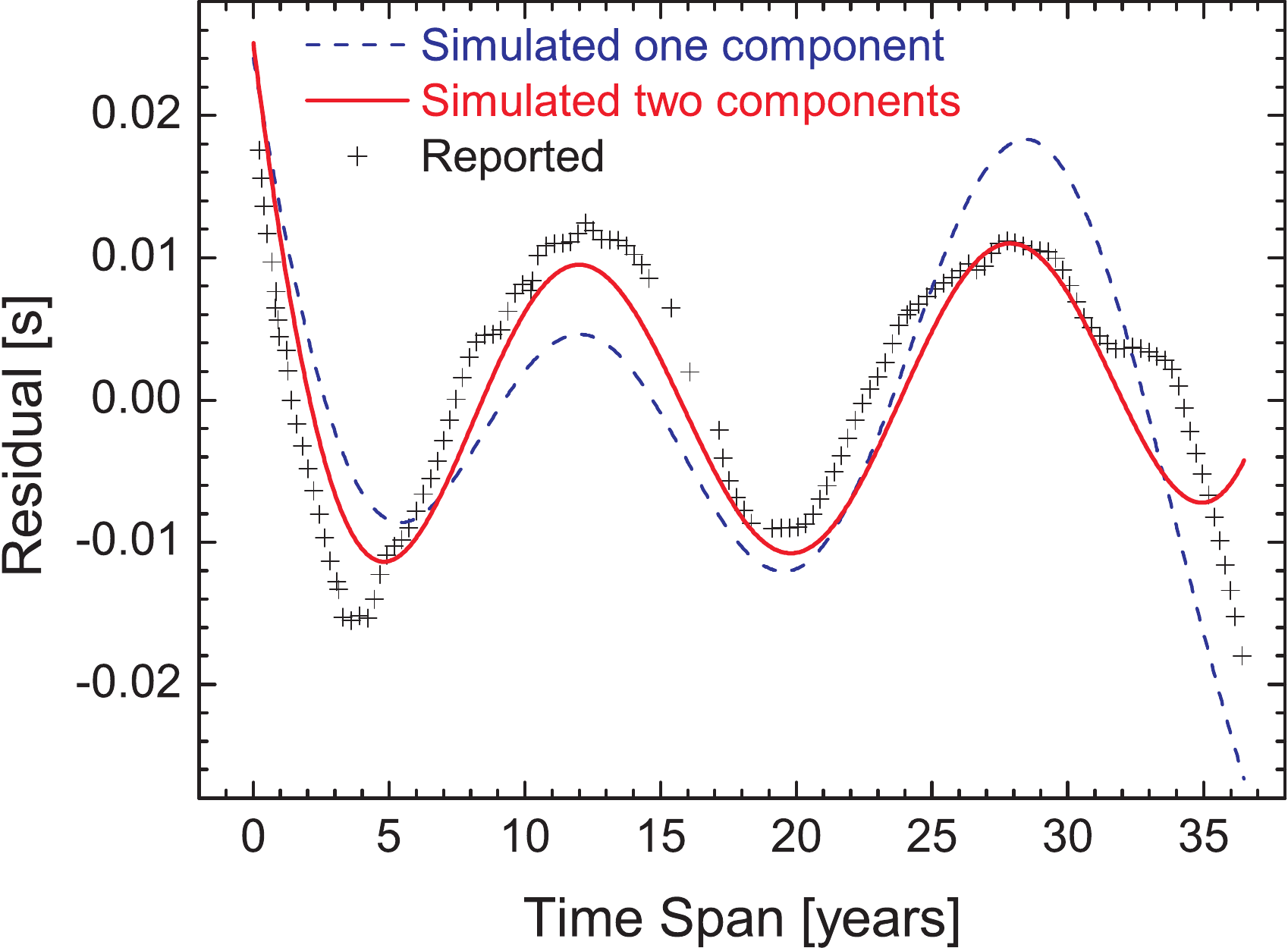}
\caption{Timing residual of PSR B0329+54. The reported timing
residual, after subtraction $\nu$ and $\dot\nu$ of the pulsar over
the $36.5$ years, is represented by cross symbols. The simulated
results with one and two components are represented by dashed and
solid lines, respectively. The model parameters are identical with
that for the $\ddot\nu$ simulation, shown in Figure {\ref{Fig:1}}.}
\label{Fig:2}
\end{figure}

The timing residual, after subtraction of the pulsar's $\nu$ and
$\dot\nu$ over $36.5$ years for PSR B0329+54, is also simulated with
exactly the same model parameters used for modeling $\ddot\nu$. In
the simulation, the following steps are taken:

(i) We get the model-predicted TOAs with $\Delta T_{\rm
int}=10^6~{\rm s}$ using Equation (\ref{phase integrate}) over
$36.5~{\rm yr}$, with the same model parameters used for modeling
$\ddot\nu$.

(ii) By fitting the TOA set $\{\Phi(t_i)\}$ to
\begin{equation}\label{phase2}
\Phi(t) =\Phi_0 + \nu_0 (t-t_0) + \frac{1}{2}\dot\nu_0 (t-t_0)^2,
\end{equation}
we get $\Phi_0$, $\nu_0$ and $\dot\nu_0$.

(iii) Then the timing residual after the subtraction of $\nu$ and
$\dot\nu$ can be obtained by
\begin{equation}\label{residual}
T_{\rm res}(t_i)=\frac{\Phi(t_i)-(\Phi_0 + \nu_0 (t_i-t_0) +
\frac{1}{2}\dot\nu_0 (t_i-t_0)^2)}{\nu_0}.
\end{equation}

In Figure \ref{Fig:2}, we plot the reported timing residual
(from Figure 3 of H2010) with cross symbols and the
simulated results for one oscillation component and two components
with dashed and solid lines, respectively. Note that the
simulated results are not fits of the models to the reported timing
residuals; the model parameters are set after many rounds of trials and comparisons with the reported data. One can see that the two-component model matches the observed
data better than the one-component model. Our model implies that the
timing residual is also caused by the magnetic field oscillation,
and the quasi-periodic structures in timing residuals have the same
origin (which is determined by Equation (\ref{braking law2})) with
those in $\ddot\nu$, $\dot\nu$, and $\nu$ variations.

In general, the two-component model describes the variation
of $\ddot\nu$ and timing residuals of PSR B0329+54 more precisely
than the one-component model. However, the oscillation component
with $50$ years period cannot be tested directly from the power
spectrum of its timing residuals, since the period is longer than
the observational data span. However, there are still some features
demonstrating its existence. For instance, the observed data are
reported about four years ago, and the two-component model predicts
that $\ddot\nu$ of the pulsar is now experiencing another switch
from positive to negative (as shown in Figure 1), which can be
tested with the latest observed data. The test could also be
conducted by applying the model to a larger set of pulsars, which
have short oscillation periods (shorter than the observed time
span), and relatively large oscillation amplitudes (so that the
swing behavior of $\ddot\nu$ could emerge; the exact criteria of $k$
depend on $\nu$, $\dot\nu$ and $T$).

\section{Simulating the Distribution of $\ddot\nu$ and its Correlation with $\tau_{\rm c}$}

\begin{figure}
\centering
\includegraphics[angle=0,scale=0.5]{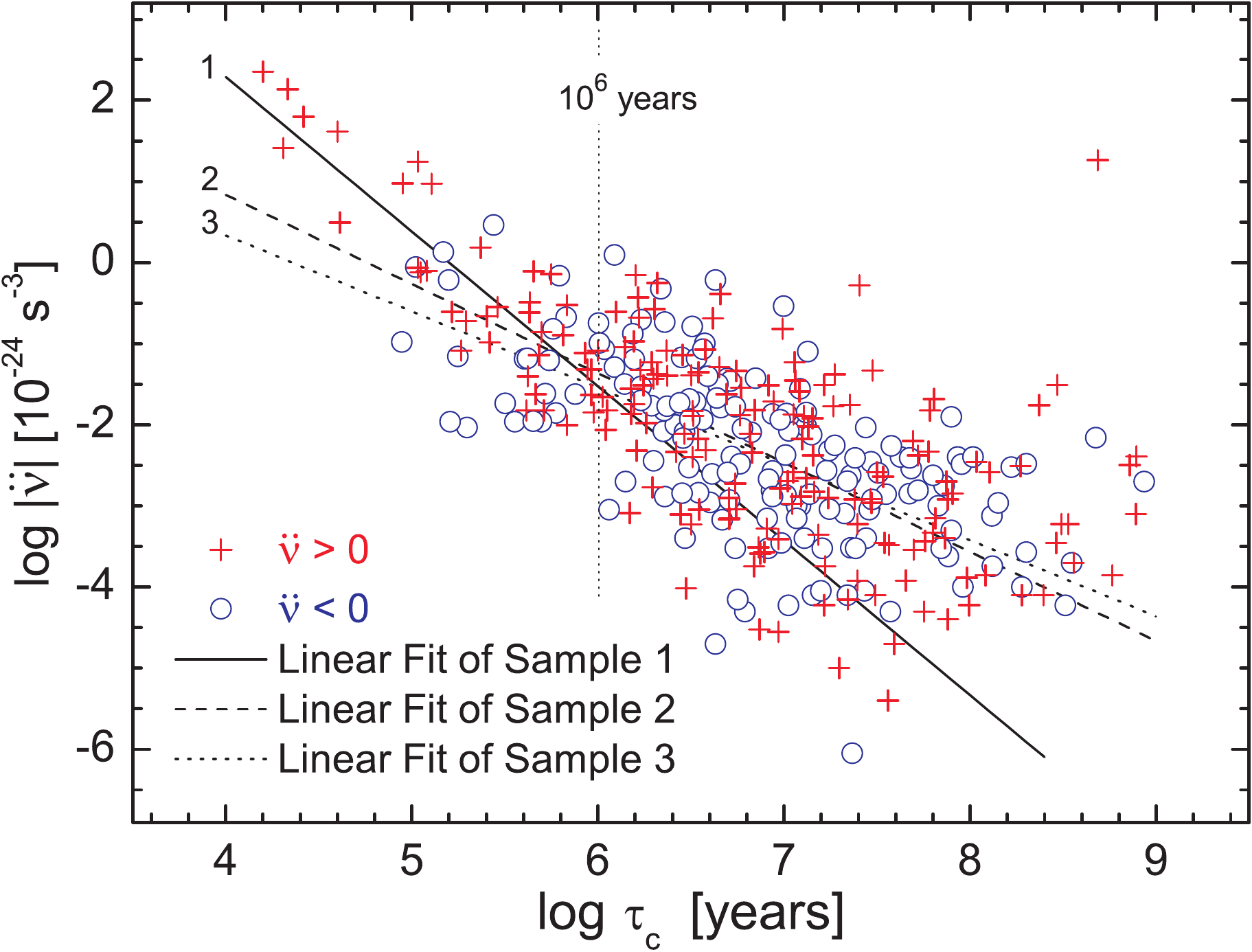}
\caption{The reported $|\ddot\nu|$ versus characteristic age
$\tau_{\rm c}$. The reported data are obtained from Table 1 of
H2010. The crosses and circles indicate $\ddot\nu>0$ and
$\ddot\nu<0$, respectively. The solid line indicates the linear fit
for $\ddot\nu>0$ and $\tau_{\rm c}<10^6~{\rm yr}$ (sample 1), the
dashed line indicates the linear fit for $\ddot\nu>0$ and $10^6~{\rm
yr}<\tau_{\rm c}<10^8~{\rm yr}$ (sample 2), and the dotted line
indicates the linear fit for $\ddot\nu<0$ and $\tau_{\rm
c}<10^8~{\rm yr}$ (sample 3). The vertical line indicates $\tau_{\rm
c}=10^6~{\rm yr}$.} \label{Fig:3}
\end{figure}

We show the measured $|\ddot\nu|$ versus $\tau_{\rm c}$ for $341$
normal radio pulsars with $\tau_{\rm c}<10^9~{\rm yr}$ in Figure
{\ref{Fig:3}} (the reported data are obtained from Table 1
of H2010). The linear fits for $\log|\ddot\nu|~[10^{-24}~{\rm
s^{-3}}]=a+b \log\tau_c~[{\rm yr}]$ are given. It is found that the
slope $b$ for $\tau_{\rm c}<10^6~{\rm yr}$ ($b_1=-2.0$) is obviously
steeper than that for $\tau_{\rm c}>10^6~{\rm yr}$ ($b_2=-1.1$) for
$\ddot\nu>0$; the latter is slightly steeper than the slope for
$\ddot\nu<0$ ($b_3=-0.94$). It was found that $b_1$ is caused by the
magnetic field decay, which dominates the field evolution for young
pulsars with $\tau_{\rm c}\lesssim10^6~{\rm yr}$ (Paper I). In this
section, based on our phenomenological model, we use the Monte Carlo
method to simulate the distributions of $\ddot\nu$ and $n$, and
their correlation with $\tau_{\rm c}$.

\begin{figure}
\centering
\includegraphics[angle=0,scale=0.5]{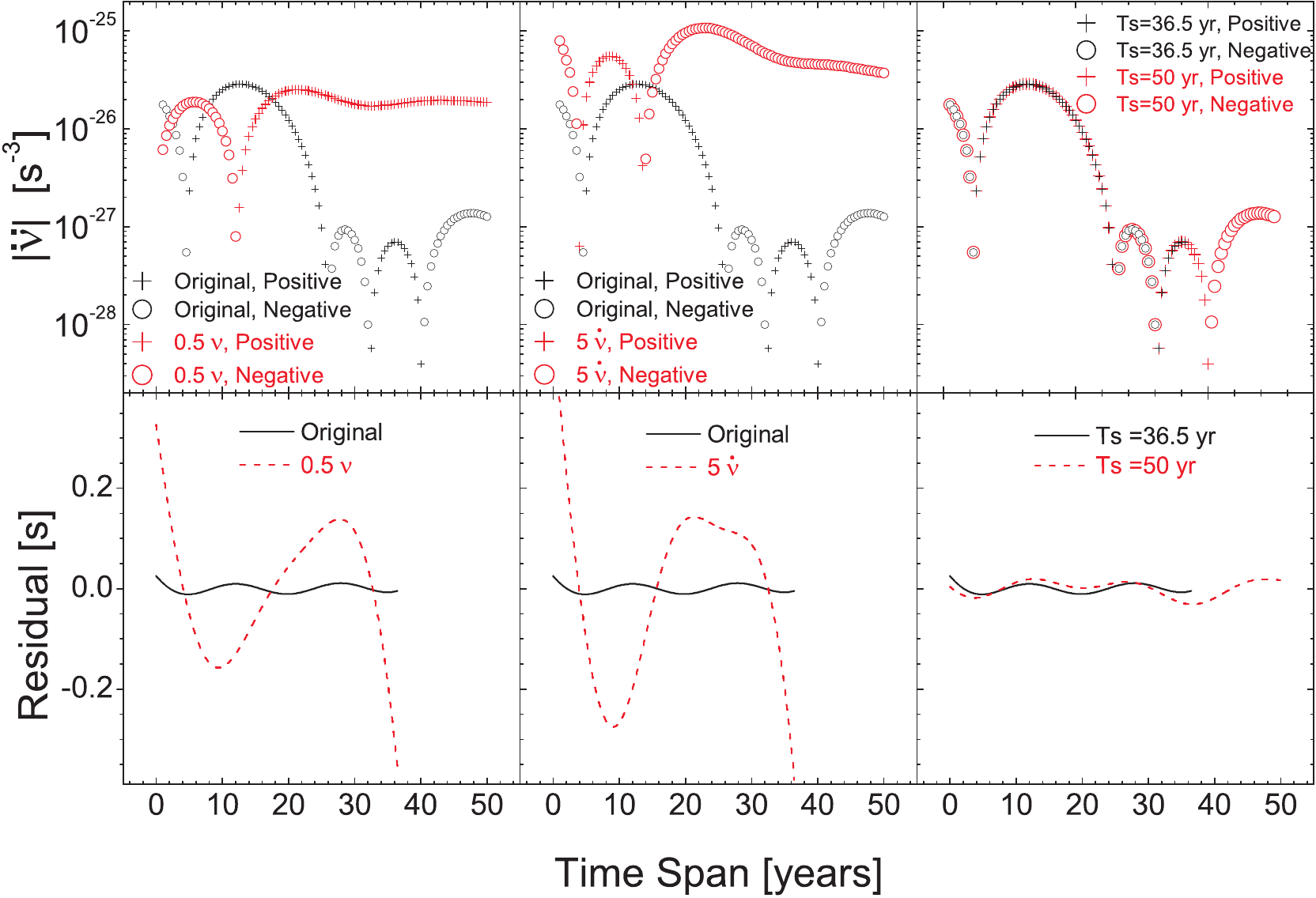}
\caption{The effects of $\nu$, $\dot\nu$ and $T_{\rm s}$ on $\ddot\nu$ and timing residuals. The results for different
$\nu$ ($\nu$ and $0.5\nu$) are shown in the left panels, the results
of different $\dot\nu$ ($\dot\nu$ and $5 \dot\nu$) are shown in
middle panels, and the results of different $T_{\rm s}$ (36.5 yr and
50 yr) are shown in left panels.} \label{Fig:r2}
\end{figure}

\subsection{Determining the Sample Space}

We firstly check the effects of the variations of $\nu$,
$\dot\nu$ and $T_{\rm s}$ on $\ddot\nu$ and timing
residuals. We still adopt the two component model of PSR B0329+54
and the same model parameters obtained above. Based on the model, we
show $|\ddot\nu|$ and timing residuals for different values of
$\nu$, $\dot\nu$ and $T_{\rm s}$ in Figure {\ref{Fig:r2}}. In the
left panels of Figure {\ref{Fig:r2}}, one can see that both
$|\ddot\nu|$ and timing residuals for $0.5\nu$ ($\nu$ is the
pulsar's reported value, and other parameters are fixed to their
reported values) are apparently larger than that of $\nu$. The
results are similar for the case of $5\dot\nu$ in the middle panels.
In the right panels, one can see that $\ddot\nu$ and the timing
residuals have very small changes for $T_{\rm s}=36.5$ yr and $50$
yr, thus the results are not sensitive to $T_{\rm s}$. We therefore need to determine the sample space well for $\nu$ and
$\dot\nu$, but only approximately for $T_{\rm s}$.

\begin{figure}
\centering
\includegraphics[scale=0.5]{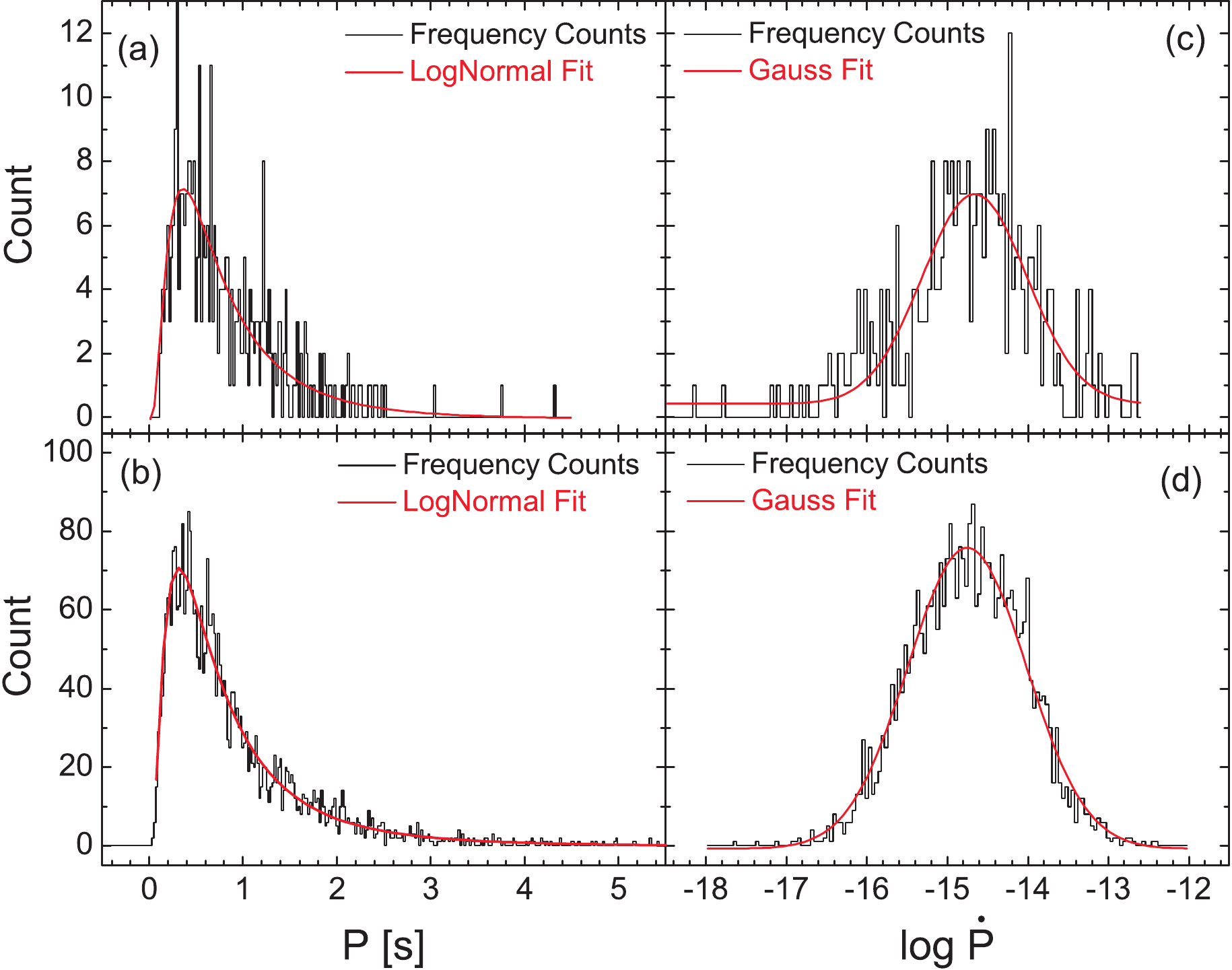}
\caption{Distributions and fits for $\nu$ and $\log\dot\nu$. (a)
Distribution of $\nu$ for the reported sample and their lognormal
fit. (b) Distribution of $\nu$ for the simulated sample and their
lognormal fit; the selection effect of ``death line'' is included
for the sample space. (c) Distribution of $\log\dot\nu$ for the
reported sample and their Gauss fit. (d) Distribution of
$\log\dot\nu$ for the simulated sample and their Gauss fit; the
selection effect of ``death line'' is included for the sample
space.} \label{Fig:4}
\end{figure}

The distribution of $\nu$ can be well described by a lognormal
distribution, as shown in Figure {\ref{Fig:4}}(a). The best-fit
parameter set is $(x_{\rm c}, \sigma)=(0.66, 0.78)$, where $x_{\rm
c}$ and $\sigma$ are the mean and standard deviation, respectively.
In the simulation, we choose the same distribution as the sample
space for $\nu$ with $(x_{\rm c}, \sigma)=(0.67, 0.86)$. We show
about 3500 sample outcomes in Figure {\ref{Fig:4}}(b), and their
best-fit parameter set $(0.66, 0.85)$ approximately equals the set
of the reported sample after the selection effect of the ``death
line'' is included. The theoretical ``death line'' we adopt here is
$7\log B-13\log P=78$ (Chen \& Ruderman 1993), where the dipole
magnetic field $B$ is in units of Gauss and period $P$ is in units
of seconds. For different simulated samples, the best-fit parameter
sets have small fluctuations that can be ignored.

The distribution of $\log\dot\nu$ can be well described by a
Gaussian distribution, as shown in Figure {\ref{Fig:4}}(c). The
best-fit parameter set is $(\mu, \sigma)=(-14.66, 1.3)$, in which
$\mu$ and $\sigma$ are the mean and standard deviation,
respectively. Similarly, the parameter set $(-14.8, 0.7)$ is adopted
for the sample space, and the best-fit parameter set $(-14.75,
1.47)$ for the 3500 sample outcomes approximately equals the set of
the reported sample after the selection effect of the ``death line''
is considered, as shown in Figure {\ref{Fig:4}}(d).

\begin{figure}
\centering
\includegraphics[angle=0,scale=0.5]{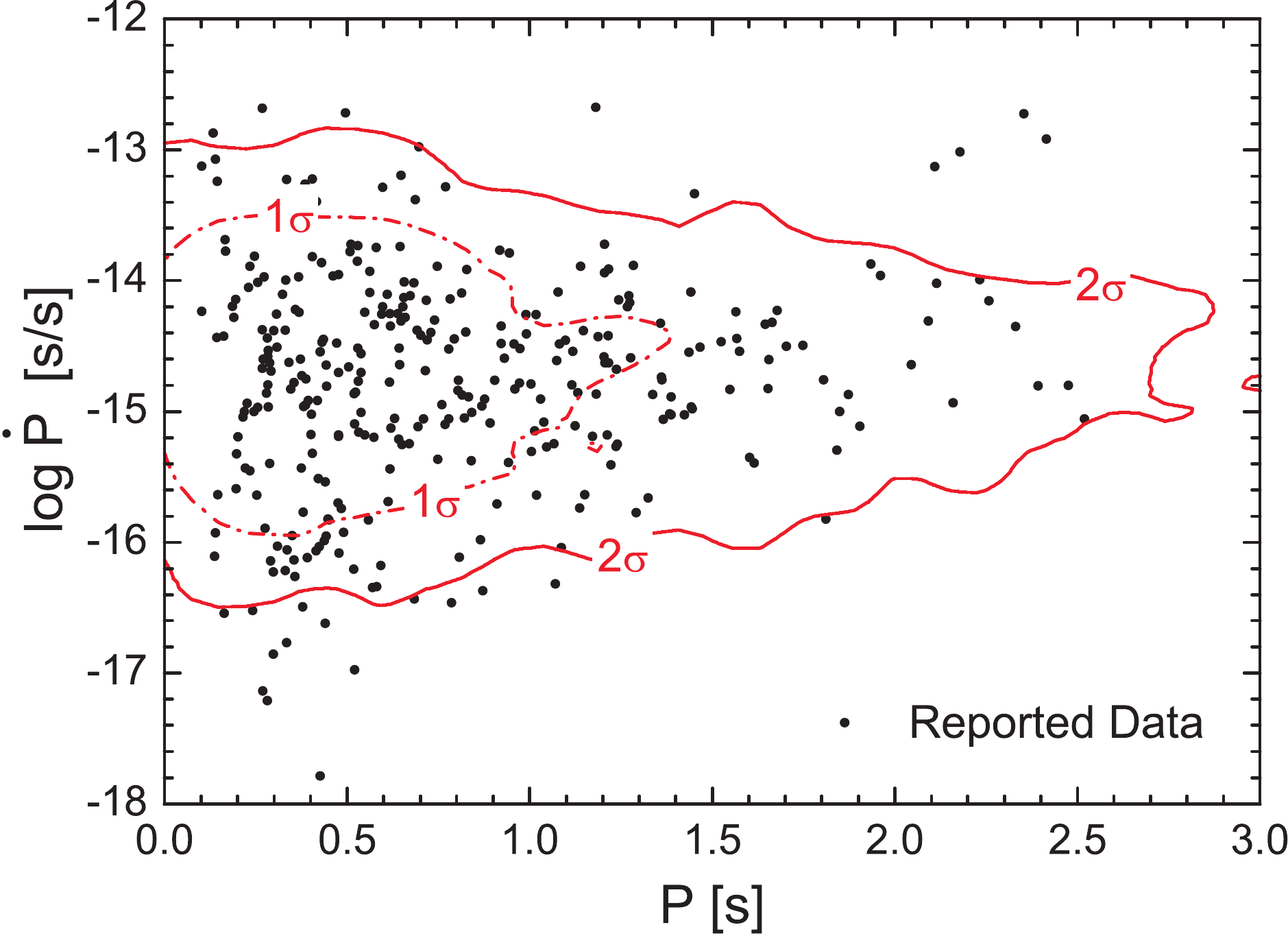}
\caption{The reported data points and contour lines for the
$\sim3500$ simulated sample outcomes in the $P-\log\dot P$ diagram.
The $1\sigma$ line (dash-dotted line) indicates the area covering
$\sim 68 \%$ sample outcomes, and $2\sigma$ line (solid line)
indicates the area of $\sim 95\%$ sample outcomes.} \label{Fig:5}
\end{figure}

We plot the $P-\log\dot P$ diagram for the reported sample and the
contour lines of the $\sim3500$ simulated sample outcomes in Figure
\ref{Fig:5}, in which the period $P=1/\nu$. One can see that the
simulated sample agrees with the reported sample very well, about
$93\%$ of the reported data are covered by the $2\sigma$ area of the
simulated data and $62\%$ of the reported data are covered by the
$1\sigma$ area.

\begin{figure}
\centering
\includegraphics[angle=0,scale=0.5]{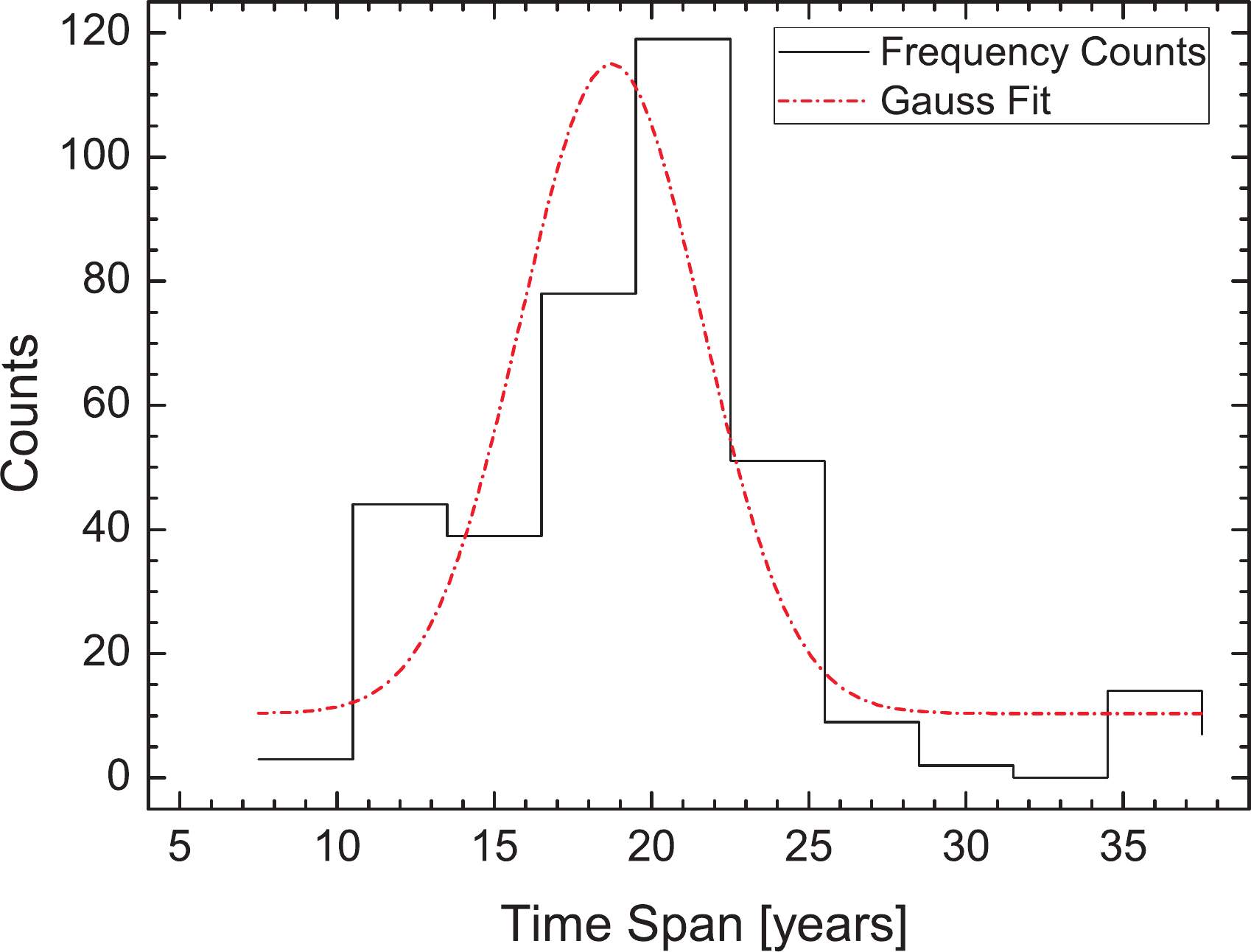}
\caption{Histogram of the time spans of H2010 observations and its
Gaussian fit.} \label{Fig:6}
\end{figure}

\begin{figure}
\centering
\includegraphics[angle=0,scale=0.5]{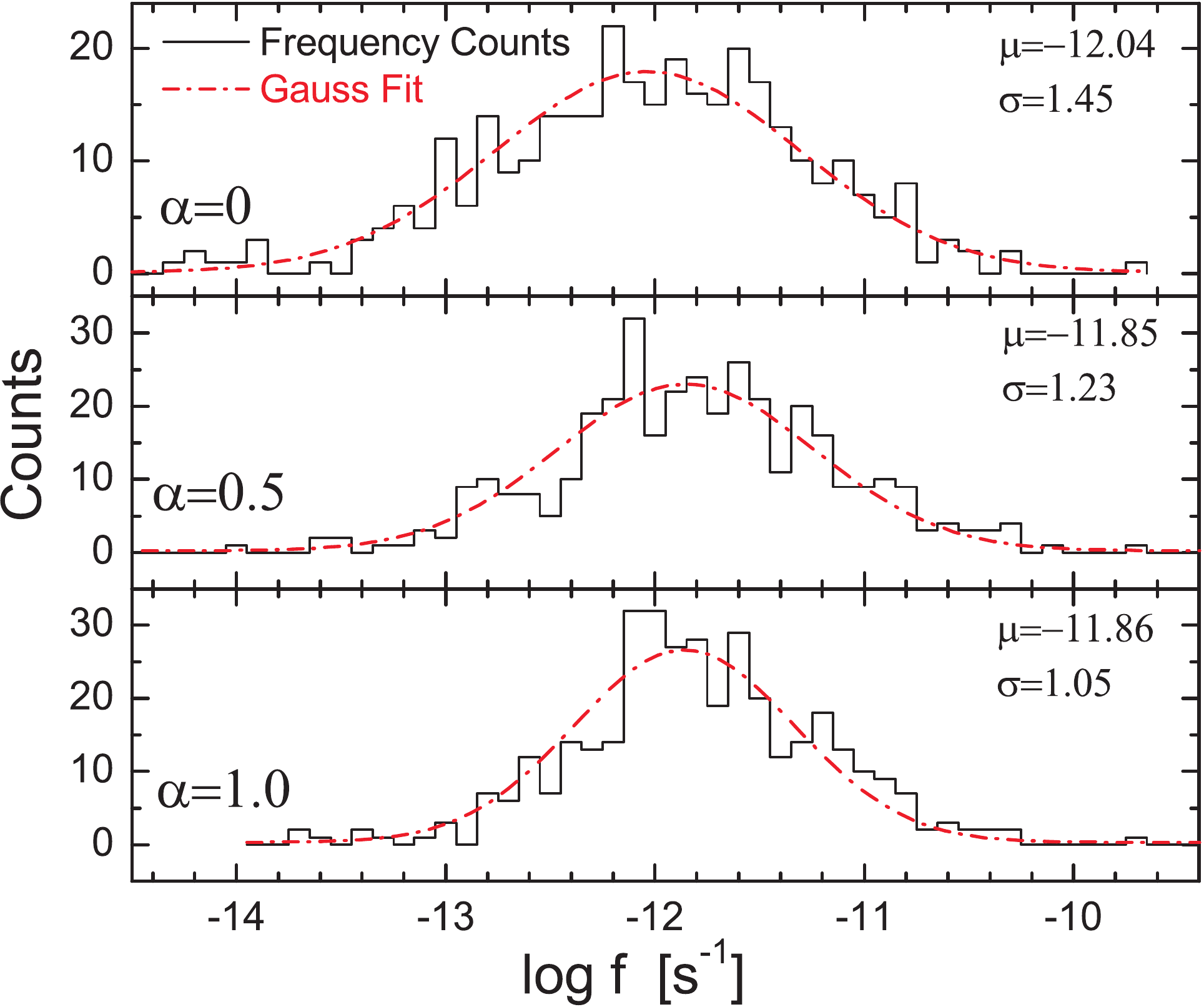}
\caption{Histograms of the oscillation parameter $f$ and their
Gaussian fits for different long-term magnetic field decay index
($\alpha$).} \label{Fig:9}
\end{figure}

The histogram of the time spans of observations $T_{\rm s}$ (H2010)
and its Gaussian fit are shown in Figure {\ref{Fig:6}}. The best-fit
parameter set $(\mu, \sigma)$ is $(18.7~{\rm yr}, 5.76~{\rm yr})$,
which determines the sample space for the upper limit of Equation
(\ref{phase integrate}). Though the distribution of $T_{\rm
s}$ is poorly modelled by the Gaussian, it is still good enough for
the simulation, since $\ddot\nu$ and timing residuals are not
sensitive to $T_{\rm s}$, as shown in the right panels of Figure
\ref{Fig:r2}.

From Equation (\ref{braking law2}), we obtained the analytic
approximation (in Paper I) for $\ddot\nu$
\begin{equation}\label{apprddot}
\ddot\nu\simeq -2\dot\nu(\alpha/t \pm f),
\end{equation}
where $f=2\pi k/T$ represents the magnitude of the oscillation term.
Thus, both parameters $k$ and $T$ are important. In our previous
work (Paper I), we get $\log f$ for all pulsars in the sample of
H2010 by the following steps: (a) we set $\alpha=0$, $0.5$ or $1.0$;
(b) for a certain value of $\alpha$, we can get $\eta$ by fitting
the data of young pulsars with $\tau_c<2\times10^6~{\rm yr}$ to
Equation (15) in Paper I, where $\eta$ is defined as
$\eta=(3.3\times 10^{19}/B_0)^{1/\alpha}2\alpha /t_0 $; (c) then
$f$ can be obtained by substituting $\nu$, $\dot\nu$ and $\ddot\nu$
of each pulsar into Equation (14) (in Paper I). One can see that the
distribution of $\log f$ shows a single peak for a certain $\alpha$
(see Figure 8 in Paper I). We show the distributions for different
values of $\alpha$ and their Gaussian fits in Figure {\ref{Fig:9}}.
The fitted parameter set $(\mu,\sigma)$ is $(-12.04, 1.45)$,
$(-11.85, 1.23)$ and $(-11.86, 1.05)$ for $\alpha=0$, $0.5$ and
$1.0$, respectively. If we assume that the period $T$ of magnetic
field oscillations is a constant, the sample space parameters of $k$
can be obtained from $k=f T/2\pi$.

\begin{figure}
\centering
\includegraphics[angle=0,scale=0.7]{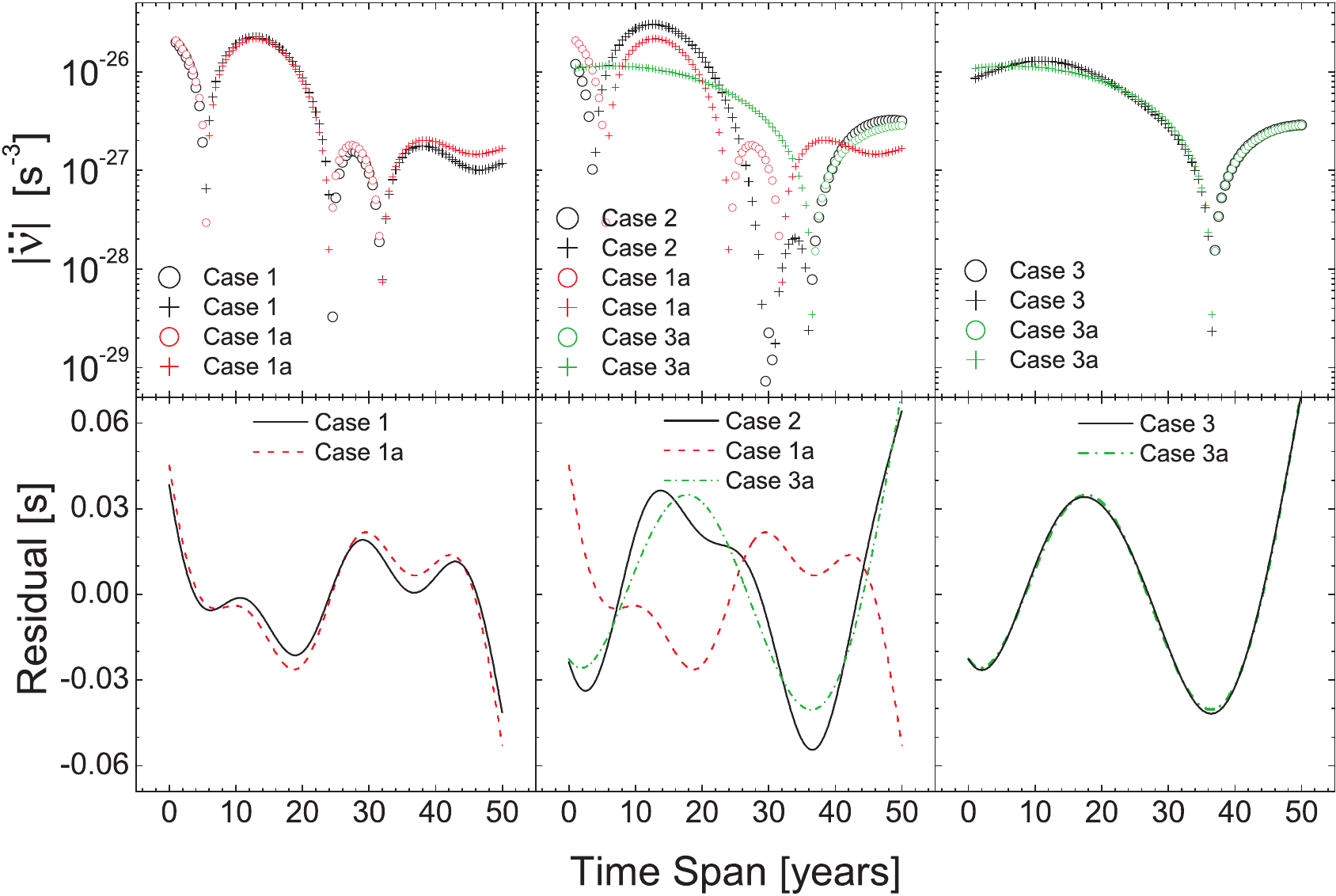}
\caption{Comparisons of $\ddot\nu$ and timing residuals with one
oscillation component and two oscillation components. The model of
B0329+54 is still adopted. Case 1a: one oscillation component with
$k=3\times10^{-4}$ and $T=16.8~{\rm yr}$. Case 1: two oscillation
components with $k_1=3\times10^{-4}$, $T_1=16.8~{\rm yr}$ (the short
period is the dominating component), and $k_2=3\times10^{-5}$,
$T_2=50~{\rm yr}$. Case 2: two oscillation components with
$k_1=3\times10^{-4}$, $T_1=16.8~{\rm yr}$, and $k_2=3\times10^{-4}$,
$T_2=50~{\rm yr}$ (no dominating component). Case 3a: one
oscillation component with $k=3\times10^{-4}$ and $T=50~{\rm yr}$.
Case 3: two oscillation components with $k_1=3\times10^{-5}$,
$T_1=16.8~{\rm yr}$ and $k_2=3\times10^{-4}$, $T_2=50~{\rm yr}$ (the
long period is the dominating component).} \label{Fig:r3}
\end{figure}

Sometimes it might be necessary to take multiple oscillation
components, since multiple peaks are often seen in the power spectra
of the timing residuals of many pulsars (H2010). However, to our
knowledge there is not any statistical data on the numbers of the
oscillation components as well as their periods reported up to now
in the literature. For simplicity, here we assume that there is
always a dominating oscillation component (Paper I), which mainly
determines the variations of $\ddot\nu$ and the timing residuals. In
Figure \ref{Fig:r3}, we compare $\ddot\nu$ and the timing residuals
between the two component and the one component case. It is found
that if one of the components dominates, the two component model can
be well approximated by the one component model which has the same
$k$ and $T$ as the dominating one. However, if the two components
have comparable $k$, the approximation is no longer valid (as shown
in the middle panels), and some uncertainties may be introduced,
which we have to live with currently. In addition, since $T$ is also
not well known to date, we will try several different values for it
in the following Monte Carlo simulations.

\subsection{Results of Monte Carlo Simulations}

One can draw a set of $\nu$, $\dot\nu$, $T_{\rm s}$, $T$, $k$, and
$\phi$ from the above sample space. The sample of the phase $\phi$
of the field oscillation follows a uniform random distribution in
the range of $0$ to $2\pi$. With these quantities and a
corresponding start time $t_0$, we can obtain a rotation phase set
$\{\Phi(t_i)\}$ using Equation (\ref{phase integrate}). In the
calculation, the time interval of TOAs is also assumed as a
constant, i.e. $\Delta T_{\rm int}=10^6~{\rm s}$. Then the
``averaged'' values of $\nu$, $\dot\nu$ and $\ddot\nu$ can be
obtained by fitting $\{\Phi(t_i)\}$ to Equation (\ref{phase}). Hence
one has its $|\ddot\nu|$ and $\tau_{\rm c}$. Repeat this procedure
for $N$ times, we will have $N$ data points in the
$|\ddot\nu|$-$\tau_{\rm c}$ diagram. In Table 1 we summarize all the
model parameters and results of simulations.

\subsubsection{Effects of power-law decay index}

\begin{figure}
\centering
\includegraphics[scale=0.6]{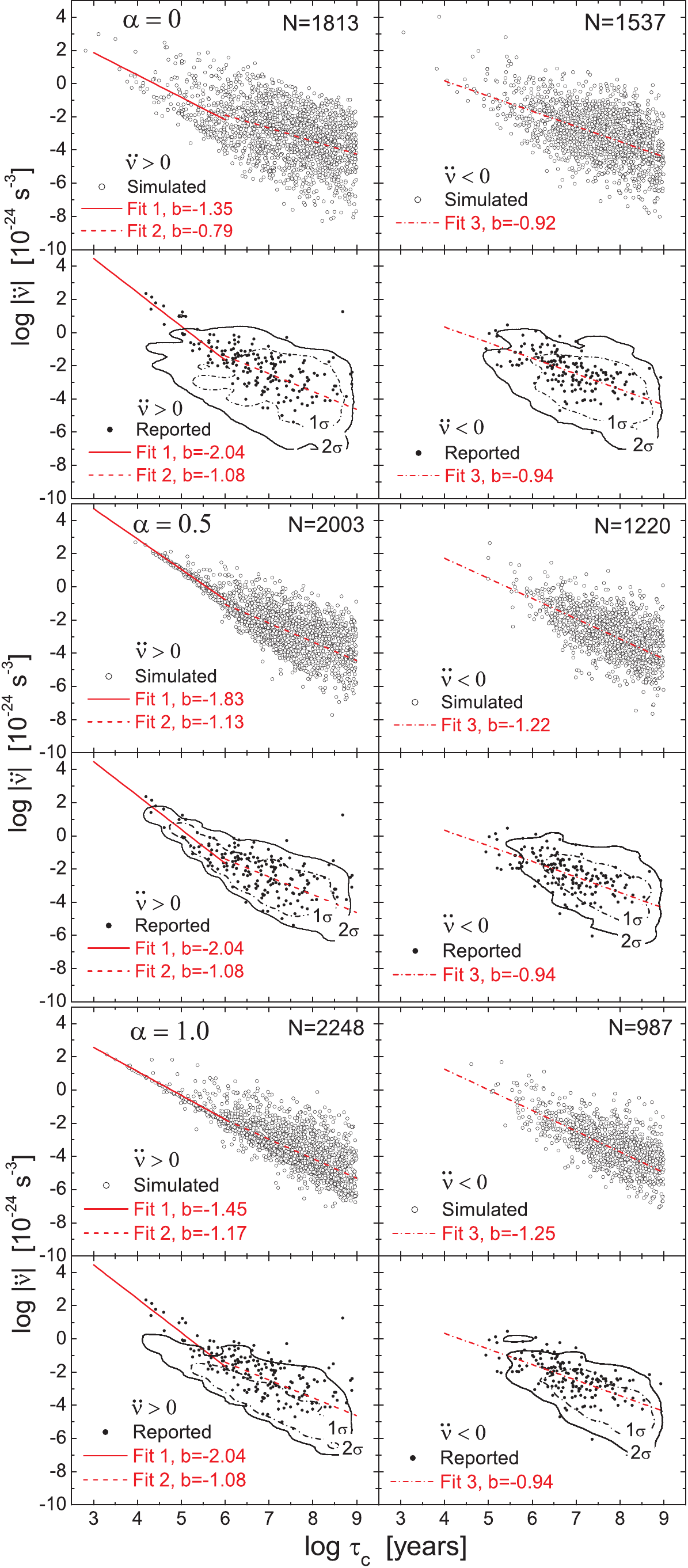}
\caption{Simulations of the $|\ddot\nu|$-$\tau_{\rm c}$ distribution
in the cases of no long-term magnetic field decay ($\alpha=0$, the
upper four panels), moderate long term magnetic field decay
($\alpha=0.5$, the middle four panels), and strong long term
magnetic field decay ($\alpha=1$, the bottom four panels). Linear
fit 1: for $\ddot\nu>0$ with $\tau_{\rm c}<10^6~{\rm yr}$ (young
pulsars); Linear fit 2: for $\ddot\nu>0$ with $10^6~{\rm
yr}<\tau_{\rm c}<10^8~{\rm yr}$ (old pulsars); Linear fit 3: for
$\ddot\nu<0$ with $\tau_{\rm c}<10^8~{\rm yr}$. The best-fit slopes
($b$) for all simulations and reported data are labelled in these
corresponding panels.} \label{Fig:7}
\end{figure}

\emph{Case I: no long-term decay, i.e. $\alpha=0$.} For this case,
we assume $B_0=3.2\times 10^{19}\sqrt{-\dot\nu/\nu^3}$ and
$t_0=-\nu/2\dot\nu$, and the oscillation period $T=20~{\rm yr}$. We
plot the simulated results in the upper four panels of Figure
\ref{Fig:7}. The number of the total data points is $N_{\rm
total}=3350$ (the number is not fixed for each simulation, due to
the selection effect of ``death line''), in which the numbers of
positive and negative $\ddot\nu$ are $N_{\rm p}=1813$ and $N_{\rm
n}=1537$, respectively. The distribution contours and the reported
data are also shown for $\ddot\nu>0$ and $\ddot\nu<0$, respectively.
For $\ddot\nu>0$ about $94\%$ and $63\%$ of the reported data are
covered by the $2\sigma$ and $1\sigma$ areas of the simulated data,
respectively; similarly, for $\ddot\nu<0$ about $94\%$ and $73\%$ of
the reported data are covered by the $2\sigma$ and the $1\sigma$
areas, respectively. However, the steep slope for the young pulsars
with $\tau_{\rm c}<10^6~{\rm yr}$ cannot be well reproduced.

\emph{Case II: power-law decay with $\alpha=0.5$.} For this case, we
assume $t_0=H_1/B_0^2$, where $H_1=3.075\times10^{36}~({\rm G^2s})$
is obtained by the best-fit for the reported young pulsars with
$\tau_{\rm c}>10^6~{\rm yr}$ and $\ddot\nu>0$ (Paper I), and
$T=20~{\rm yr}$. We plot the simulated results in the middle four
panels of Figure \ref{Fig:7}. $N_{\rm total}=3223$, in which $N_{\rm
p}=2003$ and $N_{\rm n}=1220$, respectively. $N_{\rm p}/N_{\rm
total}$ ($\sim 62\%$) is larger than the reported $\sim54\%$. For
$\ddot\nu>0$ about $95\%$ and $74\%$ of the reported data are
covered by the $2\sigma$ and $1\sigma$ areas of the simulated data,
respectively; similarly, for $\ddot\nu<0$ about $88\%$ and $63\%$ of
the reported data are covered by the $2\sigma$ and the $1\sigma$
areas, respectively. Notably, a steeper slope for the young pulsars
with $\tau_{\rm c}<10^6~{\rm yr}$ can almost be reproduced.

\emph{Case III: power-law decay with $\alpha=1.0$.} For this case,
we assume $t_0=H_2/B_0^2$, where $H_2=5.16\times10^{24}~({\rm G~s})$
is also obtained by the best-fit for the reported young pulsars with
$\tau_{\rm c}>10^6~{\rm yr}$ and $\ddot\nu>0$ (Paper I), and
$T=20~{\rm yr}$. We plot the simulated results in the bottom four
panels of Figure \ref{Fig:7}. $N_{\rm total}=3235$, in which $N_{\rm
p}=2248$ and $N_{\rm n}=987$, respectively. $N_{\rm p}/N_{\rm
total}$ is larger than $69\%$. For $\ddot\nu>0$, about $81\%$ of the
reported data are covered by the $2\sigma$ area of the simulated
data, but only $34\%$ of the reported data are covered by $1\sigma$
area; for $\ddot\nu<0$ about $90\%$ and $55\%$ of the reported data
are covered by the $2\sigma$ and the $1\sigma$ areas, respectively.
However, the slope for the young pulsars with $\tau_{\rm
c}<10^6~{\rm yr}$ is still not steep enough.

In conclusion, it is found that $\alpha=0.5$ is favored by the
reported data.

\begin{deluxetable}{lcccccccccccccccccccc}
\tabletypesize{\scriptsize} \tablecaption{Summary for all the
simulated results. The first row lists the reported data in our
sample. The numbers in the left and rights parts of brackets
correspond to $\ddot\nu>0$ and $\ddot\nu<0$, respectively.}

\tablewidth{0pt} \tablehead{\colhead{}& \multicolumn{4}{c}{Model
Parameters}&  & \multicolumn{4}{c}{Results}& & \colhead{Note}& \\

\cline{2-5} & \cline{5-10}

\colhead{$ $}& $\alpha$ & $\log f$ & $k~(10^{-4})$ & $T$ (yr) &
$(N_{\rm p}, N_{\rm n})$ & \colhead{$b_1$}& \colhead{$b_2$} &
\colhead{$b_3$} & \colhead{$1\sigma~($\%$)$}&
\colhead{$2\sigma$~($\%$)}}

\startdata
$~$ & -- & -- & --  & -- &~ (183, 158) & -2.04 & -1.08 & -0.94 & -- & -- & Fig {\ref{Fig:3}}\\

\hline

$~$ & 0.0 & -12.04 & 0.92 & 20 &~ (1813, 1537) & -1.35 & -0.79 & -0.92 & (94, 94) & (63, 73) & Fig {\ref{Fig:7}}\\

$~$ & 0.5 & -11.85 & 1.42 & 20 &~ (2003, 1220) & -1.83 & -1.13 & -1.22 & (95, 88) & (74, 63) & Fig {\ref{Fig:7}}\\

$~$ & 1.0 & -11.86 & 1.38 & 20 &~ (2248, ~987) & -1.45 & -1.17 & -1.25 & (95, 88) & (74, 63) & Fig {\ref{Fig:7}}\\

\hline

$~$ & 0.0 & -12.04 & 0.23 & 5 &~ (2061, 1099) & -1.09 & -1.75 & -1.82 & -- & -- & Fig {\ref{Fig:10}}\\

$~$ & 0.5 & -11.85 & 0.35 & 5 &~ (2669, ~611) & -1.93 & -1.99 & -1.82 & -- & -- & Fig {\ref{Fig:10}}\\

$~$ & 1.0 & -11.86 & 0.35 & 5 &~ (2580, ~386) & -1.49 & -1.31 & -1.26 & -- & -- & Fig {\ref{Fig:10}}\\

\hline

$~$ & 0.0 & -12.04 & 4.58 & 100 &~ (1598, 1510) & -1.41 & -0.79 & -0.96 & -- & -- & Fig {\ref{Fig:10}}\\

$~$ & 0.5 & -11.85 & 7.09 & 100 &~ (2140, 1303) & -1.71 & -1.12 & -1.29 & -- & -- & Fig {\ref{Fig:10}}\\

$~$ & 1.0 & -11.86 & 6.93 & 100 &~ (1948, 1459) & -1.15 & -0.85 & -0.95 & -- & -- & Fig {\ref{Fig:10}}\\

\hline

$~$ & 0.5 & -11.85 & 70.9  & $10^3$ &~ (2039, 1438) & -1.20 & -1.03 & -1.24 & -- & -- & Fig {\ref{Fig:12}}\\

$~$ & 0.5 & -11.85 & 709.0 & $10^4$ &~ (2306, 1463) & -1.25 & -1.21 & -1.07 & -- & -- & Fig {\ref{Fig:12}}\\

$~$ & 0.5 & -11.85 & 7090  & $10^5$ &~ (2581, ~954) & -1.54 & -1.21 & -1.04 & -- & -- & Fig {\ref{Fig:12}}\\

\hline

$~$ & 0.5 & -11.35 & 6.73  & $30$ &~ (1950, 1681) & -1.72 & -0.84 & -1.03 & (93, 90) & (69, 66) & Fig {\ref{Fig:11}}\\

\enddata
\label{Tab:1}
\end{deluxetable}

\subsubsection{Effects of Oscillation Period}

\begin{figure}
\centering
\includegraphics[scale=0.6]{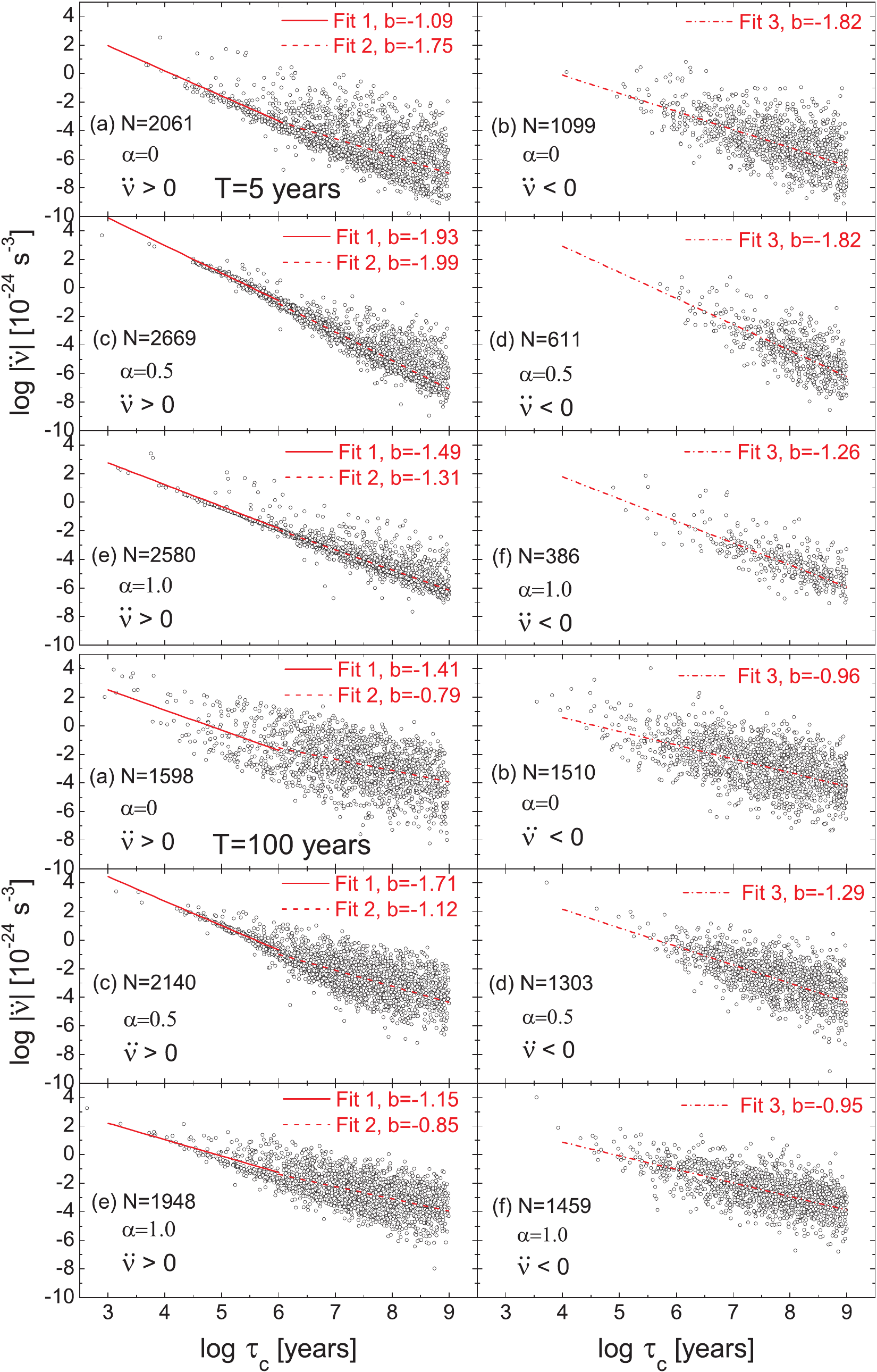}
\caption{Simulations of the $|\ddot\nu|$-$\tau_{\rm c}$ distribution
for different combinations of long-term magnetic field decay index
($\alpha$) and the period $T$, as marked in each panel. Fit 1: for
$\ddot\nu>0$ with $\tau_{\rm c}<10^6~{\rm yr}$ (young pulsars); Fit
2: for $\ddot\nu>0$ with $10^6~{\rm yr}<\tau_{\rm c}<10^8~{\rm yr}$
(old pulsars); Fit 3: for $\ddot\nu<0$ with $\tau_{\rm c}<10^8~{\rm
yr}$. $N$ is the number of data points.} \label{Fig:10}
\end{figure}

\emph{The case of $T=5~{\rm yr}$.} We keep all parameters the same
as those in the above subsection, except that the oscillation period
is changed to $5~{\rm yr}$. The main results are shown in the upper
six panels of Figure \ref{Fig:10}. One can see that $N_{\rm p}\gg
N_{\rm n}$. It can be inferred that the oscillation has impacts
mainly on older pulsars, since $\ddot\nu<0$ appears mostly in the
area with larger $\tau_{\rm c}$. The slopes (i.e. $b$)
for the young pulsars with $\ddot\nu>0$ are too flat, but the slopes
for the old $\ddot\nu>0$ and $\ddot\nu<0$ are too steep. In
addition, one can see that there is a crowded area of data points
along the lower boundary for $\ddot\nu>0$. The crowded area is
caused by the underestimation for $|\ddot\nu|$, because the
``averaging'' effect is strong when the oscillation period is much
shorter than the observation time span. However, there is no such
crowded area in the reported data, which indicates that the period
$T=5~{\rm yr}$ is too short for most pulsars in the sample.
Simulations show that there is no obvious crowded area when the mean
value of period is longer than thirty years, i.e. $T\gtrsim 30~{\rm
yr}$, which is actually beyond the $2\sigma$ range of the sample
space for the observation time span, and thus the ``averaging''
effect does not dominate the reported $\ddot\nu$.

\emph{The case of $T=100~{\rm yr}$.} We keep all parameters the same
but the oscillation period is changed to $100~{\rm yr}$. We show the
simulated results in the lower six panels of Figure {\ref{Fig:10}}.
It is found that $N_{\rm p}\approx N_{\rm n}$
for $\alpha=0$. But for the cases of $\alpha=0.5$ and $\alpha=1.0$,
$N_{\rm p}\gg N_{\rm n}$ ($N_{\rm p}/N_{\rm
total}=62\%$ and $57\%$, respectively). As expected by the above
analysis, there is not a clear crowded area for this case. The
steep slope ($b$) for the young pulsars with $\ddot\nu>0$ can only
be reproduced by $\alpha=0.5$; this suggests again that
$\alpha=0.5$ dominates the long-term magnetic field
decay for young pulsars with $\tau_{\rm c}<10^6~{\rm yr}$.

\begin{figure}
\centering
\includegraphics[scale=0.6]{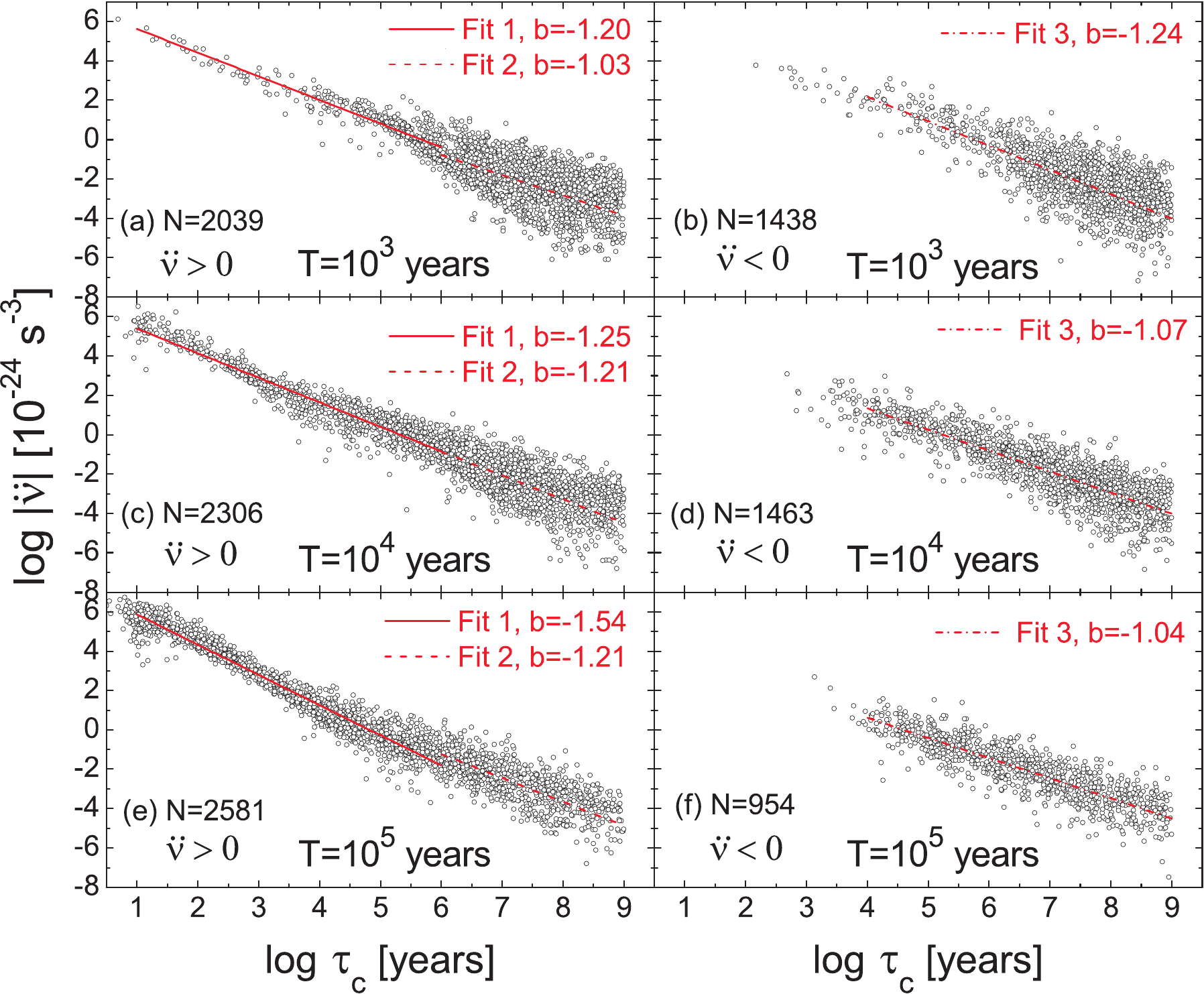}
\caption{Simulations of the $|\ddot\nu|$-$\tau_{\rm c}$ distribution
in the cases of moderate long term magnetic field decay
($\alpha=0.5$) for long oscillation period, $T=10^3$, $10^4$ and
$10^5~{\rm yr}$. Linear fit 1: for $\ddot\nu>0$ with $\tau_{\rm
c}<10^6~{\rm yr}$; Linear fit 2: for $\ddot\nu>0$ with $10^6~{\rm
yr}<\tau_{\rm c}<10^8~{\rm yr}$; Linear fit 3: for $\ddot\nu<0$ with
$\tau_{\rm c}<10^8~{\rm yr}$.} \label{Fig:12}
\end{figure}

\emph{The cases of $T=10^3, 10^4$ and $10^5~{\rm yr}$.} The results
of simulations for $T=10^3, 10^4$ and $10^5~{\rm yr}$ are shown in
Figure \ref{Fig:12}. One can see that there are many simulated data
points spread over from $\tau_c=10^5$ to $10~{\rm yr}$ as shown in
the left panels for $\ddot\nu>0$, which is apparently different from
reported data. For the reported data and simulated data of
$T\lesssim100~{\rm yr}$, the overall shape of data points is a
triangle like; as the period increases from $T=10^3$ to $10^5~{\rm
yr}$, the overall shape of data points gradually becomes a band
like. This is due to the oscillation parameter $k\gtrsim1$ for
$T=10^4~{\rm yr}$ (since $f$ is fixed and $k=fT/2\pi$), and such a
large oscillation magnitude will deviate the $\dot\nu$ from its
initial value significantly, which is inconsistent with the
observational facts that $\dot\nu$ does not change significantly.
Thus the simulations can give a constraint for the upper limit of
oscillation period, $T\lesssim10^3~{\rm yr}$.

In conclusion, $T$ has an influence on the distribution density and
the overall shape of simulated data points when $f$ is a constant.
By comparing the distribution density and the overall shape with the
reported data, we give a rough constraint for the oscillation
period, $30~{\rm yr}\lesssim T \lesssim1000~{\rm yr}$.

\begin{figure}
\centering
\includegraphics[scale=0.4]{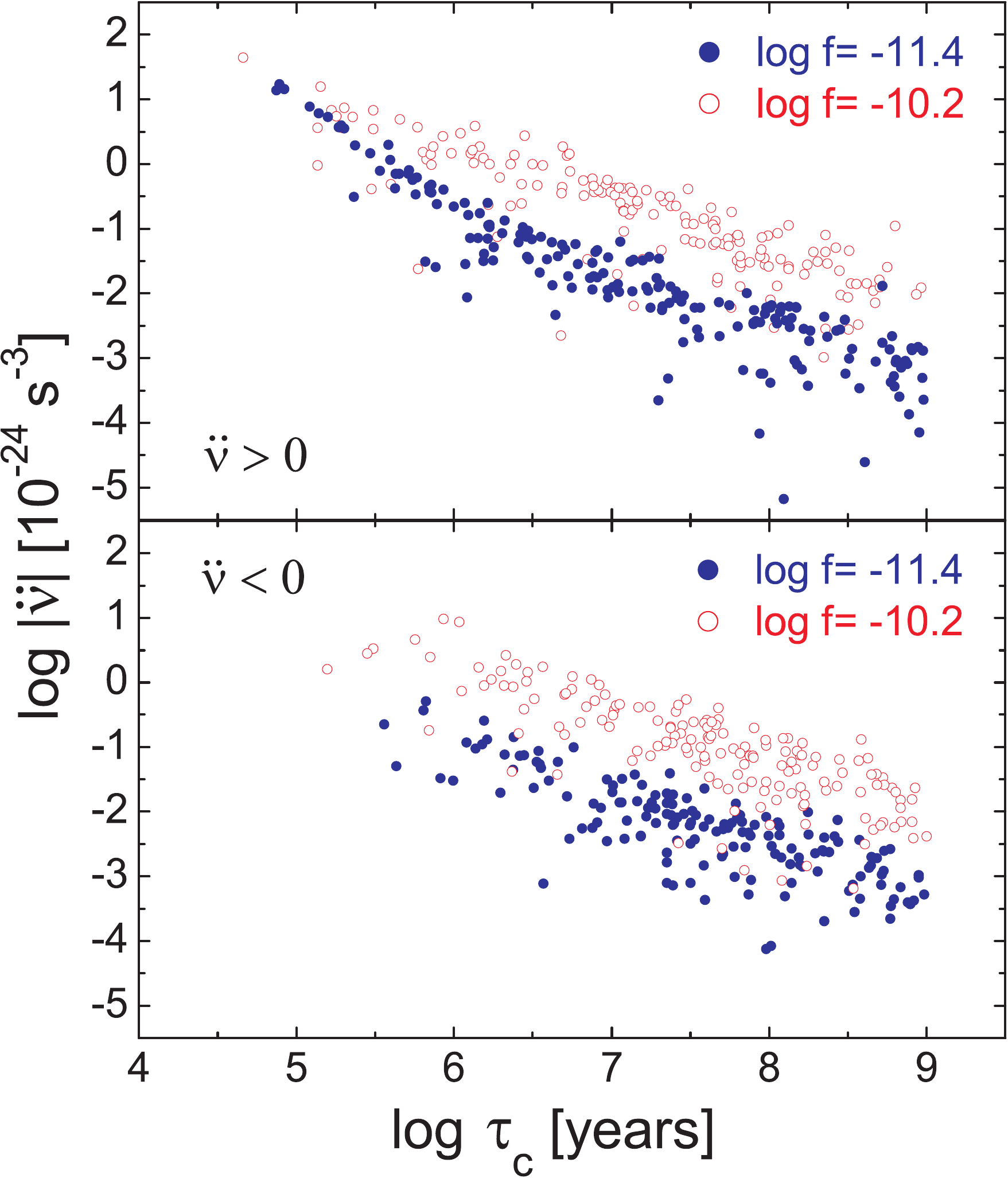}
\caption{Simulations of the $|\ddot\nu|$-$\tau_{\rm c}$ distribution
for $\alpha=0.5$, $T=30~{\rm yr}$. The results with parameter set
$(-11.35, 0.01)$ and $(-10.2, 0.01)$ for $\log f$ are shown by
unfilled circles and filled circles, respectively.} \label{Fig:13}
\end{figure}

\subsubsection{Effects of Oscillation Amplitude}

We assume the oscillation amplitude parameter set for $\log f$ is
$(\mu, \sigma)=(-11.35, 0.01)$ and $(-10.2, 0.01)$, respectively.
The simulated results for $\alpha=0.5$ and $T=30~{\rm yr}$ are shown
in Figure \ref{Fig:13}. One can obtain two conclusions from the
figure: (a) a larger $\log f$ makes an upper distribution envelop
higher (with larger $|\ddot\nu|$, as predicted by Equation (11), and
(b) the lower distribution envelop shows a steeper slope for the
segment of $\tau_{\rm c}\lesssim10^6~{\rm yr}$ for $\ddot\nu>0$.

$N_{\rm p}\approx N_{\rm n}$ is an important constraint for the
model. It is found that if $\log
f\gtrsim -11.4$ (i.e. $k\gtrsim 6\times10^{-4}$ for $T=30~{\rm
yr}$), $N_{\rm p}\approx N_{\rm n}$ for the case of $\alpha=0.5$
(e.g. we perform a simulation with $(\mu=-10.3, \sigma=1)$ and
$T=30~{\rm yr}$ and get $3494$ outcomes, in which $N_{\rm p}=1849$
and $N_{\rm p}/N_{\rm total}\simeq 52.9\%$); however, if $\log
f\lesssim-11.4$, we will obtain
$N_{\rm p}\gg N_{\rm n}$. The existence of the lower limit is mainly
due to the competition between the magnetic field long-term decay
and the short-term oscillation, as predicted by Equation
(\ref{apprddot}). However, it is worth to note that the lower limit
is larger than the analytical result $\log f\simeq-11.85$, as shown
in Figure {\ref{Fig:9}}. This is caused by the ``averaging'' effect
that induced an underestimation for $|\ddot\nu|$. Meanwhile, it is
also found that $|\ddot\nu|$ will be larger than the
$2\sigma$ range of the reported data if $\log f\gtrsim -10.2$ (i.e.
$k\gtrsim 0.01$ for $T=30~{\rm yr}$). In conclusion, the
upper and lower bounds of $\log f$ (for $\alpha=0.5$) can be
obtained by using the conditions of $N_{\rm p}\approx N_{\rm n}$ and
the upper boundary of reported data: $-11.4\lesssim\log
f\lesssim-10.2$.

\begin{figure}
\centering
\includegraphics[scale=0.6]{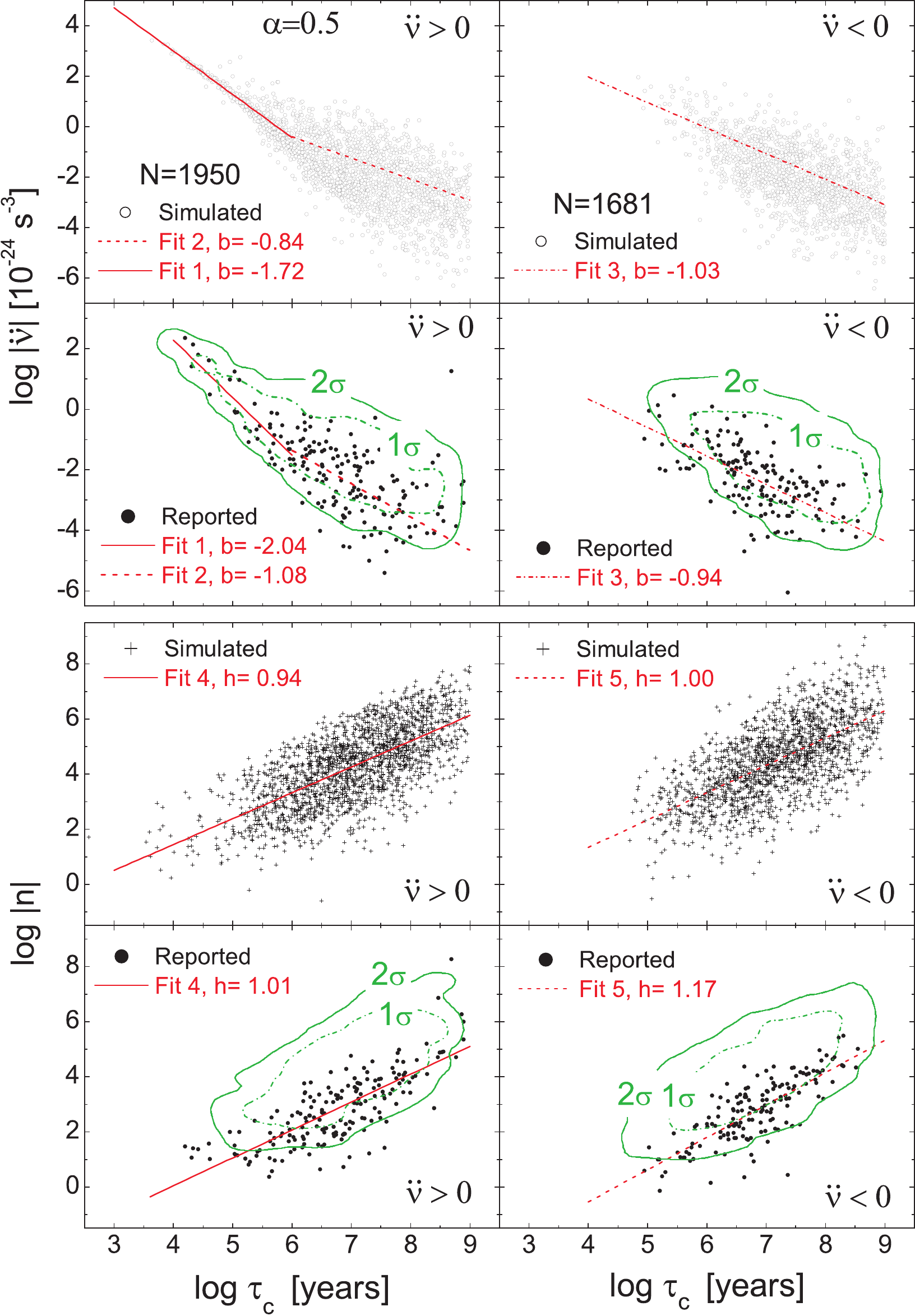}
\caption{Simulations of the $|\ddot\nu|$-$\tau_{\rm c}$ distribution
(upper panels) in the cases of moderate long term magnetic field
decay ($\alpha=0.5$) and correlations of $n$ with $\tau_c$ (bottom
panels). Linear fit 1: for $\ddot\nu>0$ with $\tau_{\rm c}<10^6~{\rm
yr}$; Linear fit 2: for $\ddot\nu>0$ with $10^6~{\rm yr}<\tau_{\rm
c}<10^8~{\rm yr}$; Linear fit 3: for $\ddot\nu<0$ with $\tau_{\rm
c}<10^8~{\rm yr}$; Linear fit 4: for $n>0$; Linear fit 5: for $n<0$.
The best-fit slopes ($b$ and $h$) for all simulations and reported
data are labelled in these corresponding panels.} \label{Fig:11}
\end{figure}

Based on the constraints for $T$ and $\log f$ and many similar
simulations as described above, we find the best parameters are:
$\alpha=0.5$, $T=30~{\rm yr}$ and $(\mu=-11.35,\sigma=1.0)$ for
$\log f$. We show the simulated results with these parameters in
Figure {\ref{Fig:11}}, in which $N_{\rm total}=3606$, $N_{\rm
p}=1950$ and $N_{\rm n}=1681$. $N_{\rm p}/N_{\rm total}\approx
54\%$. For $\ddot\nu>0$ about $93.4\%$ and $69.4\%$ of the reported
data are covered by the $2\sigma$ and $1\sigma$ areas of the
simulated data, respectively; similarly, for $\ddot\nu<0$ about
$90.0\%$ and $65.8\%$ of the reported data are covered by the
$2\sigma$ and the $1\sigma$ areas, respectively. Though the slope
($b=-1.81$) for the young pulsars with $\tau_{\rm c}<10^6~{\rm yr}$
is still slightly too flat, the three slopes are generally
consistent with slopes of reported data. Compared with the case of
$T=20~{\rm yr}$, the $2\sigma$ area overlaps with the reported data
points better.

In the bottom four panels of Figure {\ref{Fig:11}}, we compare the
observed and simulated correlations between $\tau_c$ for $n>0$ (left
panels) and $n<0$ (right panels), respectively; the general trends
of the data are also reproduced. The linear fits for
$\log|n|~[10^{-24}~{\rm s^{-3}}]=d+h \log\tau_c~[{\rm yr}]$ are
given, and for both the observed data and simulated data,
$h\simeq1$. For $n>0$ about $84.2\%$ and $59.5\%$ of the reported
data are covered by the $2\sigma$ and $1\sigma$ areas of the
simulated data, respectively; similarly, for $n<0$ about $93.0\%$
and $56.9\%$ of the reported data are covered by the $2\sigma$ and
the $1\sigma$ areas, respectively. However, one can see that the
simulated $|n|$ are systematically larger than the reported results.
This situation can be improved by setting a smaller $f$, which
however will result in $N_{\rm p}\gg N_{\rm n}$.

\subsubsection{The Two-dimensional Kolmogorov-Smirnov Test}

Here we perform the two-dimensional Kolmogorov-Smirnov
(2DKS) test to reexamine the distributions of the simulated results
using the KS2D
package\footnote{http://www.astro.washington.edu/users/yoachim/code.php}.
Our purpose is to test the consistency of the distributions of
reported data and the simulated data in Figure \ref{Fig:7}, and we
show the returned probabilities in Table \ref{Tab:2}. If the
returned probability is greater than 0.2, then it is a sign that you can treat
them as drawn from the same distribution. One can see that the
simulated results, for all values of $\alpha$, are apparently
rejected by the test. However, some of the main features of the
distributions can be reproduced by the model, as we discussed above.
On the other hand, the 2DKS test indicates that $\alpha=0.5$ is
still relatively better than the others.

The failures to the 2DKS tests mean
that our model is too simple, and the discrepancy is mainly caused
by the larger $|\ddot\nu|$ given by the model based on the sample
space. The possible reasons are: (1) the magnetic field of old
pulsars have no long-term decay; (2) the median value of the
magnetic inclination angle is apparently smaller than $\pi/2$, i.e.
$\theta\ll \pi/2$, since a smaller $\theta$ corresponds to a longer
$t$, and thus a smaller $\ddot\nu$, as predicted by Equation
({\ref{apprddot}}); (3) we assume all the pulsars have the same $k$
and $T$, and with only one oscillation component; and (4) As argued in H2010, the timing noise in some young pulsars is dominated by
``glitch recovery", which cannot be modelled by the present model and thus should cause some discrepancies from our model predictions.

\begin{table}[!h]
\tabcolsep 0pt \caption{The returned probabilities of 2DKS test for
simulated data with reported data. If the probability is greater
than 0.2, then them can be treated as drawn from the same
distribution.} \vspace*{-12pt}
\begin{center}
\def\temptablewidth{0.5\textwidth}
{\rule{\temptablewidth}{1pt}}
\begin{tabular*}{\temptablewidth}{@{\extracolsep{\fill}}ccccccc}
Database & $ P(\ddot\nu >0)$ & $P(\ddot\nu <0)$ \\
\hline
        $\alpha=0.0$    & $9.63\times 10^{-9}$ & $1.49\times 10^{-12}$  \\
        $\alpha=0.5$    & $3.09\times 10^{-6}$ & $3.14\times 10^{-9}$   \\
        $\alpha=1.0$    & $1.29\times 10^{-7}$ & $5.24\times 10^{-12}$  \\
       \end{tabular*}
       \label{Tab:2}
       {\rule{\temptablewidth}{1pt}}
       \end{center}
       \end{table}

\section{Summary and Discussion}

In this work we first modeled the $\ddot\nu$ and $n$ evolutions and
applied the obtained model parameters to simulating the timing
residuals for the individual pulsar PSR B0329+54. Using a Monte
Carlo simulation method, we simulated the distributions of pulsars
in the $|\ddot\nu|-\tau_{\rm c}$ and $|n|-\tau_{\rm c}$ diagrams,
and compared the simulation results with the reported data in H2010.
Our main results are summarized as follows:

\begin{enumerate}

\item We modeled the $\ddot\nu$ evolution of pulsar PSR B0329+54
with the phenomenological model of the evolution of $B$, which
contains a long-term decay ($\alpha=0.5$) modulated by two
short-term oscillations (upper panels of Figure~\ref{Fig:1}). The
model can reproduce the main characteristics of the $|\ddot\nu|$
variation, including the swings between $\ddot\nu>0$ and
$\ddot\nu<0$.

\item For PSR B0329+54, besides a $16.8~{\rm yr}$ component as reported by
Shabanova (1995), we find that the pulsar has an another
oscillation component with period ($\sim50~{\rm yr}$) longer than
the current span of timing observations. This two component model predicts that another swing of the sign of $\ddot\nu>0$ has happened recently or in the very near future, which can be tested by analysing the recent observation data.

\item We showed that the ``averaged'' values of $n$ are different from the
instantaneous values (bottom panels of Figure~\ref{Fig:1}), and the
oscillation abruptly decays after the first period due to the
``averaging'' effect. Using these parameters obtained from modeling
the $\ddot\nu$ evolution, we simulated the timing residuals of the
pulsar (Figure~\ref{Fig:2}), which agrees with the reported
residuals (H2010) well.

\item We performed Monte Carlo simulations for the distribution of
$|\ddot\nu|$ in the $|\ddot\nu|-\tau_{\rm c}$ diagram. Our results
for different modes of magnetic field long-term decay (i.e.
$\alpha=0$, $0.5$ and $1.0$) are presented in Figures \ref{Fig:7}
and \ref{Fig:10}. It is found that the mode of $\alpha=0.5$ may
dominate the magnetic field decay for young pulsars.

\item By overlapping the $2\sigma$ areas and comparing the distribution density
and overall shape of simulated results with the reported data, we
found that the oscillation period $30~{\rm yr}\lesssim
T\lesssim1000~{\rm yr}$.

\item The observed $N_{\rm p}\approx N_{\rm n}$ can be obtained if the oscillation parameter $\log
f\gtrsim-11.4$, which is larger the analytical prediction of $\log
f\thickapprox-11.85$ (Figure \ref{Fig:9}). This is due to the
``averaging'' effect not included in our previous analytical study. The upper limit for
the oscillation parameter is $\log f\lesssim-10.2$, which is derived
from the upper boundary of the $2\sigma$ area of reported data.

\item The distribution of $n$ is also presented with the $|\ddot\nu|-\tau_{\rm c}$
diagram in Figure~\ref{Fig:11}, and the observed correlations are well reproduced. However
the simulated envelop of $|n|$ are higher than the reported data.

\end{enumerate}

In the model, there are no significant differences for the cases with
oscillation period between thirty years to few hundred years in the
simulations. However, it is pointed out that the ``averaging''
effect still has an influence on the parameters of oscillation
amplitude, i.e. on the mean value of $\log f$. Thus, an average
period about several decades years is preferred. Pons et al. (2012)
proposed a similar model of magnetic field oscillations, obtained
pulsar evolutionary tracks in $P-\dot P$ diagram, and explained the
observed braking indices of older pulsars. In their model the
magnetic field oscillations are identified as due to the Hall drift
effect in the crust of neutron stars, with a timescale of
$(10^6-10^8)\frac{10^{12}~G}{B}~{\rm yr}$ and magnitude $\delta
B/B\sim10^{-3}$. They showed that a cubic pattern would dominate the
timing residual, on the condition that the magnitude of a sinusoidal
or a random perturbation is smaller than the magnitude of the
oscillation.

We suggest that the Hall drift effect may play a role for older
pulsars; however, it is probably not a dominant mechanism for most
pulsars, since the corresponding oscillation periods are too long.
Lyne et al. (2010) showed credible evidence that timing residuals
and $\dot\nu$ are connected with changes in the pulse width.
Therefore, timing residuals are more likely caused by the changes in
a pulsar's magnetosphere with periods about $1-100~{\rm yr}$. On the
other hand, in the $|\ddot\nu|-\tau_{\rm c}$ diagram the clusters at
the old age area ($\tau_c>10^6~{\rm yr}$) are due to the fact that
the oscillation term dominates in low $|\dot\nu|$ pulsars, as we
showed in Figure \ref{Fig:10}. Thus the oscillation period as long
as $10^6~{\rm yr}$ is not necessary. Particularly for those pulsars
like PSR B0329+54, $\ddot\nu$ switches between positive values and
negative values and $\ddot\nu$ evolution and timing residuals are
coupled. These observations cannot be understand by oscillations
with period as long as million years. However, they can be well
reproduced by the model that involves magnetic field oscillations
with periods of $\sim30-100~{\rm yr}$.

\acknowledgments

We thank Meng Yu for valuable discussions. We thank Jinyuan Liao for
helps on KS2D. The anonymous referee is thanked for valuable
comments and suggestions which helped to clarify several important
points in the manuscript. SNZ acknowledges partial funding support
by 973 Program of China under grant 2009CB824800, by the National
Natural Science Foundation of China under grant Nos. 11133002 and
10725313, and by the Qianren start-up grant 292012312D1117210.

\end{document}